\documentclass[12pt]{article}

\usepackage{multibib}
\newcites{Supp}{REFERENCES}
\usepackage{amsmath}
\usepackage{amssymb}
\usepackage{amsthm}
\usepackage{mathtools}
\usepackage{graphicx}
\usepackage{xcolor}
\usepackage{natbib}
\usepackage{algorithm}
\usepackage{subcaption}
\usepackage{booktabs}
\usepackage[noend]{algpseudocode}
\usepackage[textwidth=6in,bottom=2in,top=1in,left=1in,right=1in]{geometry}
\usepackage[font=small,labelfont=bf]{caption}
\usepackage[toc,page]{appendix}
\usepackage[affil-it]{authblk}

\theoremstyle{plain}
\newtheorem{proposition}{Proposition}
\newtheorem{theorem}{Theorem}

\theoremstyle{remark}

\newtheorem{definition}{Definition}
\newtheorem{assumption}{Assumption}

\newcommand{\showrevisions}{0}
\newif\ifshowrevisions
\showrevisionstrue   

\newcommand{\markrevised}[1]{%
    \if1\showrevisions%
        \textcolor{blue}{#1}%
    \else%
        #1%
    \fi%
}

\begin{document}

\title{Continuous and Atlas-free Analysis of Brain Structural Connectivity}

\author[1]{William Consagra\thanks{Corresponding author: wconsagra@bwh.harvard.edu}}
\author[2]{Martin Cole}
\author[2]{Xing Qiu}
\author[4]{Zhengwu Zhang}

\affil[1]{Psychiatry Neuroimaging Laboratory, Harvard Medical School}
\affil[2]{Department of Biostatistics and Computational Biology, University of Rochester Medical Center}
\affil[4]{Department of Statistics and Operations Research, University of North Carolina at Chapel Hill}


\maketitle

\begin{abstract}
Brain structural networks are often represented as discrete adjacency matrices with elements summarizing the connectivity between pairs of regions of interest (ROIs). These ROIs are typically determined a-priori using a brain atlas. The choice of atlas is often arbitrary and can lead to a loss of important connectivity information at the sub-ROI level. This work introduces an atlas-free framework that overcomes these issues by modeling brain connectivity using smooth random functions.  In particular, we assume that the observed pattern of white matter fiber tract endpoints is driven by a latent random function defined over a product manifold domain. To facilitate statistical analysis of these high dimensional functional data objects, we develop a novel algorithm to construct a data-driven reduced-rank function space that offers a desirable trade-off between computational complexity and flexibility. Using real data from the Human Connectome Project, we show that our method outperforms state-of-the-art approaches that use the traditional atlas-based structural connectivity representation on a variety of connectivity analysis tasks. We further demonstrate how our method can be used to detect localized regions and connectivity patterns associated with group differences.

\end{abstract}

\noindent%
\vfill

\section{Introduction}\label{sec:introduction}
\markrevised{The \textit{structural connectivity} (SC) of the human brain refers to the pattern of anatomical connections between different brain regions formed by white matter nerve fibers, enabling communication and information transfer essential for brain function and cognition. There is great interest in understanding the variability of this connectivity \citep{durante2017nonparametric,wang2019}, both in relation to human traits such as cognition and personality \citep{zhang2019,arroyo2021,arroyo2021b}, and in the context of mental disorder and neurodegenerative disease \citep{fornito2013, park2013}. The most common way of representing the SC data is via a discrete network based model. In this formulation, the connectivity data is represented as a symmetric adjacency matrix $\boldsymbol{A} \in \mathbb{R}^{V \times V}$, where $V$ denotes the number of disjoint regions of interest (ROIs) on the brain surface. These ROIs are determined using some predefined surface parcellation known as a brain atlas \citep{desikan2006,destrieux2010}. The matrix element $\boldsymbol{A}_{ab}$ quantifies the strength of connectivity between ROIs $a$ and $b$, and is computed using some function of the number of connections between these respective regions. A variety of statistical procedures have been developed for the joint analysis of brain networks under this representation, see \cite{chung2021} for a contemporary overview.}

\par 
The reliance on the pre-specification of an atlas in the discrete ROI-based analysis framework is problematic for at least two major reasons. First, there is no consensus as to which atlas is best for brain connectivity analysis, and therefore the selection for a given application is somewhat \textit{ad-hoc}. Analyses can be sensitive to the choice of atlas \citep{zalesky2010}, resulting in different conclusions for different parcellation schemes of the same data. Second, this approach can introduce information loss, since fine-grained connectivity information on the sub-ROI level is aggregated in the construction of the adjacency matrix. 
One way to mitigate these drawbacks is to increase the number of ROIs in the parcellation. This trend can be observed in more recent atlases, e.g., 
\markrevised{\cite{glasser2016multi} with $360$ regions and \cite{schaefer2018} with $1,000$ regions}. However, increasing the number of ROIs introduces additional challenges in data analysis. The dimension of the network grows on the order of $V^2$, rendering statistical modeling and inference very challenging for even moderately large $V$. 
\par
\begin{figure}[t]
    \centering
     \includegraphics[scale=0.6]{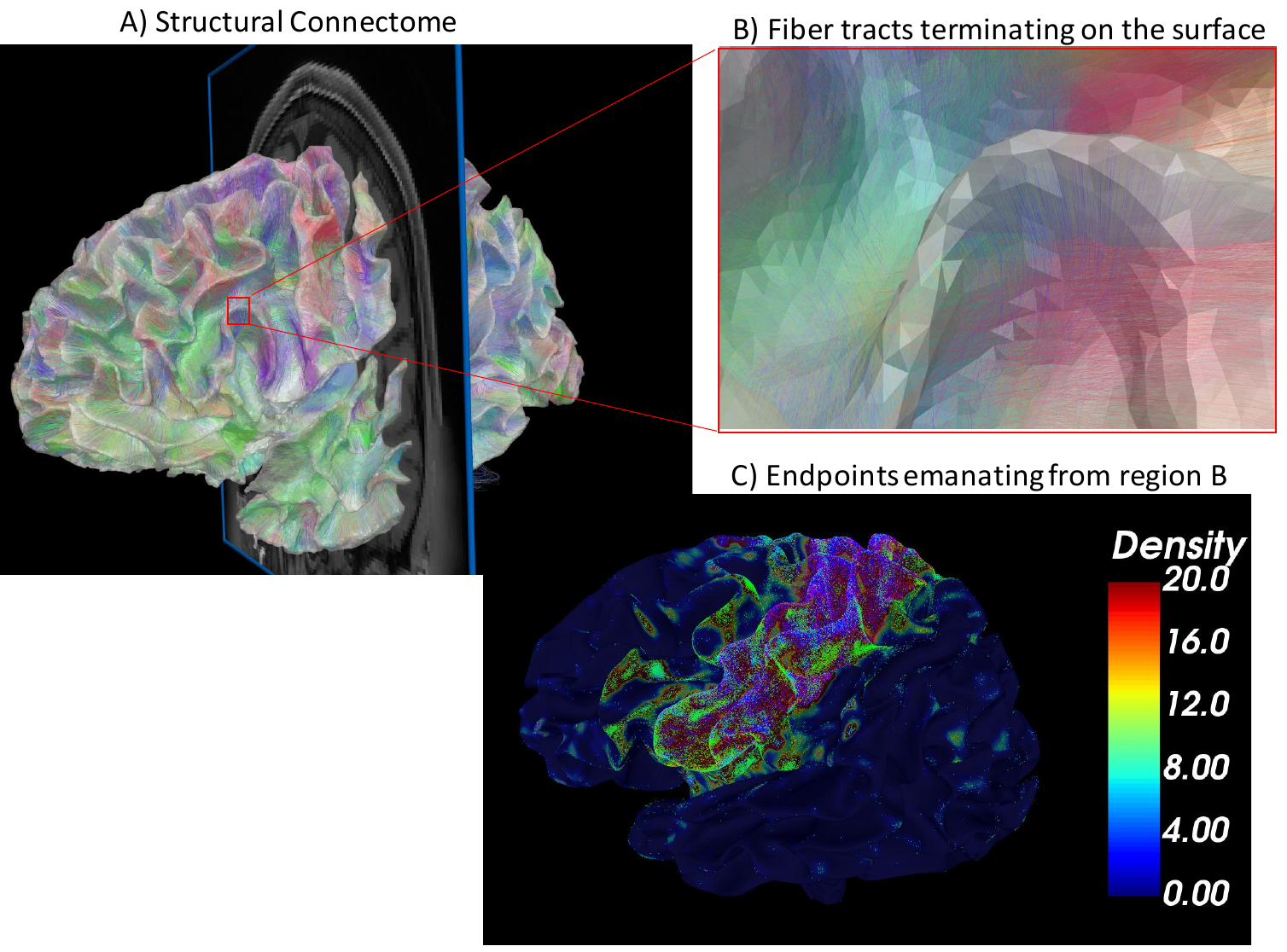}
    \caption{(A) 3D view of the cortical white surface extracted from the T1 image and the structural connectome of a randomly selected HCP subject. (B) Local surface magnification showing white matter streamlines (colored curves) ending on the surface. (C) The endpoint distribution of streamlines from the region in (B).}
    \label{fig:t1_and_surface}
\end{figure}

 
\markrevised{A collection of recent works \citep{gutman2014,moyer2017,cole2021,mansour2022} aims to address these problems by transitioning from the discrete, ROI based representation of brain connectivity to a fully continuous model.} Specifically, let $S_1$ and $S_2$ denote the left and right white surfaces of the brain. The white surface, denoted as $S_1\cup S_2$, refers to the interface between the cortical grey matter and white matter regions. Figure~\ref{fig:t1_and_surface} shows an image of the structural connectome of a randomly selected Human Connectome Project (HCP) \citep{glasser2013} subject embedded within the white surface of the brain. The endpoints of the white matter streamlines (colored curves) are points on $\left(S_1\cup S_2\right)\times\left(S_1\cup S_2\right)$. Under the continuous model, the spatial pattern of these points is assumed to be related to an unobserved continuous function on $\left(S_1\cup S_2\right)\times\left(S_1\cup S_2\right)$, which governs the strength of connectivity between any pair of points on $S_1\cup S_2$. This unobserved intensity function is referred to as the continuous (structural) connectivity, a notion first formalized in \cite{moyer2017}. 
\par
\begin{figure}[t]
    \centering
    \includegraphics[scale=0.58]{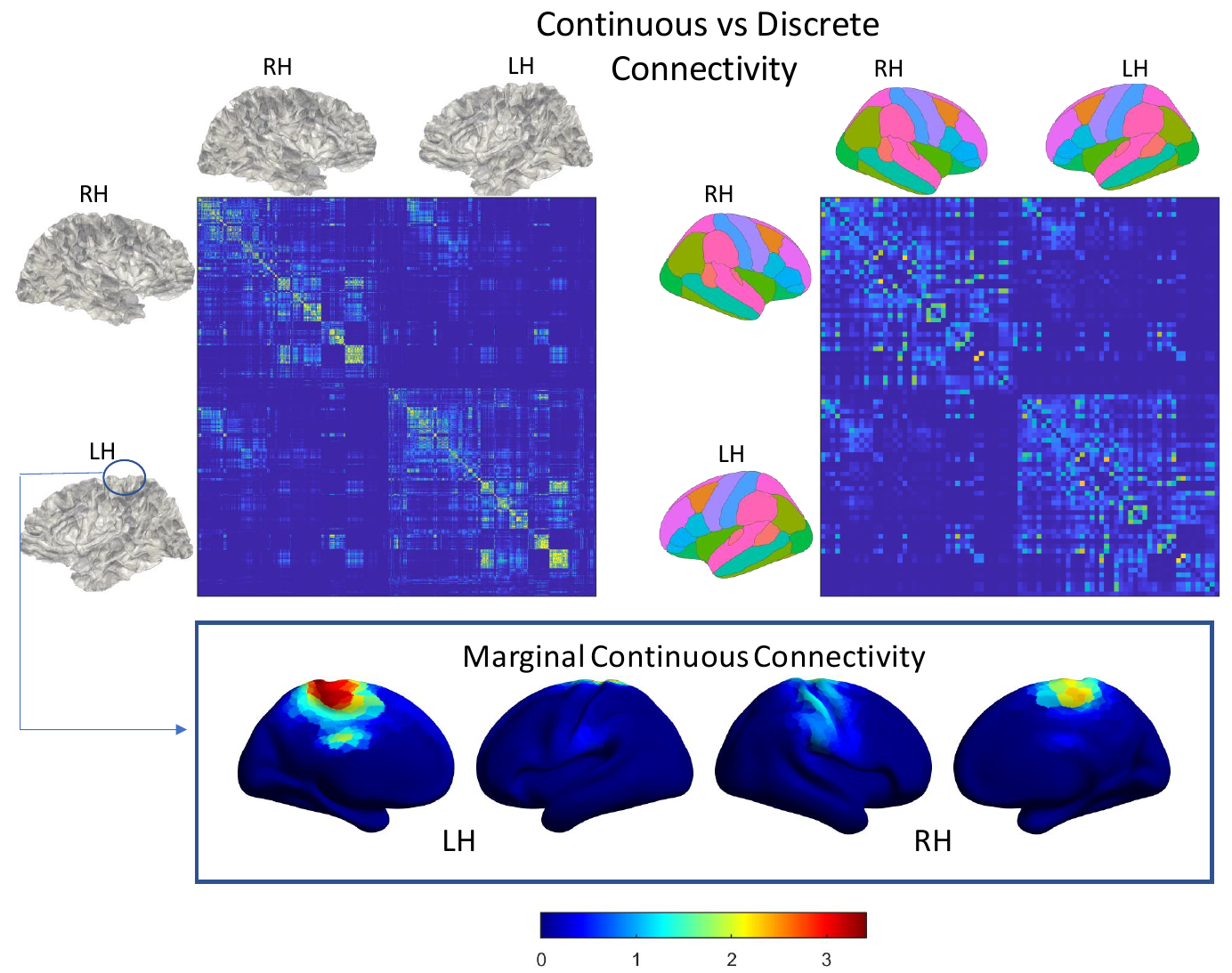}
    \caption{The top left image shows the continuous connectivity of a randomly selected HCP subject evaluated over each pair of points in a dense grid on $S_1\cup S_2$. The top right image shows the corresponding discrete connectivity matrix, under the Desikan brain atlas. The bottom panel shows the mean marginal continuous connectivity (mean of the connectivity to the circled region) in the indicated region on the left cortical surface.}
    \label{fig:connectome_representations}
\end{figure}

 
Crucially, the continuous model of connectivity does not depend on an atlas and therefore avoids the previously outlined issues which plague traditional discrete network-based approaches. Moreover, the continuous framework enables the capture of ultra-high resolution connectivity information, offering a more precise and detailed representation. For instance, the top left image in Figure~\ref{fig:connectome_representations} shows the continuous connectivity of a randomly selected HCP subject evaluated over all pairs of points in a dense grid on $S_1\cup S_2$. Compared to the corresponding discrete atlas-based connectivity matrix shown in the top right of Figure~\ref{fig:connectome_representations}, the continuous representation reveals richer and more complex patterns in the connectome. 
\par 
\markrevised{Despite the outlined advantages, the continuous approach presents additional challenges compared to the atlas-based method that have so far limited its widespread adoption. Specifically, any computation or storage requires some form of discretization of the continuous model. The current approaches discretize the continuous connectome by forming point-wise estimates over all pairs of points in a high-resolution mesh grid on $S_1\cup S_2$. For even moderately large grids, this generates enormous matrices, resulting in major computational hurdles for anything beyond simple subject-level analysis. Given this context, we make the following major contributions to advance the statistical analysis of continuous connectivity data:
\begin{enumerate}
    \item We extend the existing individual-level continuous connectivity framework \citep{gutman2014,moyer2017,cole2021,mansour2022} to a population-level framework by considering the subject-level continuous connectivity as a realization of a latent \textit{random intensity function} that governs the observed white matter streamline endpoints through a doubly stochastic point process model. Under our functional data model, we show how to perform canonical multi-subject statistical analysis tasks and propose a testing procedure to identify local subnetwork differences between groups. 
    \item We develop a novel methodology and estimation algorithm to construct a data-adaptive reduced-rank function space for efficiently representing and analyzing the  continuous connectivity. Such methodological development is required, as existing functional data analysis (FDA) approaches to point-process modeling have mainly been concerned with the 1-D case \citep{bouzas2006,muller2013,panaretos2016,wrobel2019} and have relied heavily on the spectral decomposition of the covariance function of the latent intensity process. 
    This is problematic in our case due to the curse of dimensionality. The covariance function is a constrained functional object over an 8-dimensional manifold domain, making direct estimation of this object, and hence the existing approaches, computationally infeasible.
    \item Using a large brain imaging  dataset from the HCP, we demonstrate the superiority of the continuous framework over the popular discrete framework for understanding the relationship between brain connectivity and behavioral traits.
\end{enumerate}}
While some initial findings of this study were previously presented in a brief conference paper by \cite{Consagra2022}, the current paper significantly expands and enhances that work, introducing several innovative contributions. We offer a more detailed account of the methodological framework, which includes an explanation of the construction of the reduced-rank representation space and a detailed derivation of the estimation algorithm. In response to a crucial problem in brain network analysis, a new subsection addresses the identification of brain regions with distinct interconnections across groups, including theoretical results that describe key properties of the proposed inference procedure. 
Furthermore, we include a new and extensive simulation study that examines aspects such as convergence and computational complexity of the proposed estimation algorithm and statistical power/type I error of the hypothesis testing procedure. We also significantly bolster the real data analysis by adding new comparisons, demonstrating the superiority of our method compared to several state-of-the-art competitors. This comprehensive exposition enables a broader and deeper exploration of the ideas first introduced in the conference paper.

\section{Structural Connectivity Data}\label{sec:data}
\markrevised{In this section, we introduce the HCP dataset and outline the image processing procedure used to extract the structural connectivity data.}
\subsection{Human Connectome Project}
In this work, we consider a sample of 437 female subjects from the HCP young adult cohort (\textit{https://db.humanconnectome.org}). In addition to multimodal imaging data, associated with each subject are a set of measurements related to various cognitive, physical, socioeconomic and psychiatric factors, or \textit{traits}. Many of these traits are defined and measured using tests from the NIH Toolbox for Assessment of Neurological and Behavioral Function \citep{gershon2013}, though some additional tests for measuring cognitive and emotional processing were also performed.  To understand the association between brain structural connectivity and human traits, we collected 175 traits spanning eight categories: cognition, motor, substance use, psychiatry, sense, emotion, personality and health. 

\subsection{Data Description, Image Acquisition and Processing}\label{ssec:image_processing}
Diffusion magnetic resonance imaging (dMRI) measures the local anisotropy of water molecule diffusion to infer white matter microstructure. Specifically, the local geometry of the white matter induces an anisotropy in the local water molecule diffusion, with water tending to diffuse faster along the direction of the underlying neural fiber tracts. Diffusion MRI exploits this relationship and collects measurements of spatially localized diffusion signals along many different directions, called b-vectors, to obtain a 3-dimensional picture of diffusion at each location (voxel) on a regular grid over the brain volume. Smooth local models of diffusion are fit to these measurements, e.g. the diffusion tensor \citep{Basser_diffusion} or the orientation distribution function (ODF) \citep{tuch_2004}, and are subsequently used to trace out the large-scale white matter fibers using a process called tractography \citep{basser2000}.
\par 
For each subject, we use both the dMRI and the structural T1-weighted images, the latter of which provides images with high contrast between white and grey matter regions and is therefore useful for estimating the white surface. The full imaging acquisition protocol as well as the minimal preprocessing pipeline applied to the dMRI data are given in \cite{glasser2013}, which includes susceptibility induced distortion correction and motion correction using FSL \citep{smith2004}. The cortical white surfaces $S_1$ (left) and $S_2$ (right) were estimated from the T1 image using Freesurfer and their geometry represented using a surface triangulation with $\approx 64,000$ vertices. To estimate the SC, we first estimate the local models of diffusion from the dMRI data using the approach from \cite{tournier2007} and then apply the surface enhanced tractography (SET) algorithm \citep{st2018surface} to ensure the ending points of streamlines are on $S_1\cup S_2$. 
\par 
For the joint analysis of imaging data from multiple subjects, an additional source of unwanted variability comes from misalignment due to different shapes and sizes of the brain.  To alleviate this issue, we parameterize each of the white surfaces using spherical coordinates via the surface inflation techniques from \cite{fischl1999}, and then apply a warping function estimated from aligning geometric features on the surface in Freesurfer to bring the white matter streamline endpoints to a common template space.
The final surface endpoint connectivity data consists of counts of connections between $\approx 64,000^2$ pairs of points, which presents challenges in terms of data storage. To alleviate the disk space burden, we downsampled the registered surfaces to a resolution of $4,121$ vertices using a procedure that minimizes the local metric distortion \citep{cole2021}. \markrevised{It's important to note that this downsampling was purely for storage convenience, and that the proposed method is scalable to analyze data at much higher resolutions (see Supplemental Section~\ref{sec:high_res_vs_low_res} for an example).}

\section{Statistical Framework for Continuous Connectivity}\label{sec:statistical_framework}
\markrevised{In this section, we outline our structural connectivity data generating model and describe a kernel density estimator for single subject point-wise  estimation.}
\subsection{Modeling Continuous Connectivity using Doubly Stochastic Point Processes}\label{ssec:statistical_model}
For a single subject, let $\tilde{O}=\{ (\tilde{ p}^1_1,\tilde{ p}^1_2), \cdots, (\tilde{ p}^q_{1},\tilde{ p}^q_2) \}$ denote the endpoints of $q$ streamlines connecting cortical white surfaces $\Omega = S_1 \cup S_2$ (see Figure~\ref{fig:t1_and_surface}). Since $S_i$ ($i\in \{1,2\}$) is homeomorphic to $\mathbb{S}^2$, we parameterize it using spherical coordinates.  Let $({ p}_1,{ p}_2)$ be the image of $(\tilde{ p}_1,\tilde{ p}_2)$ on $\mathbb{S}^2_1 \cup \mathbb{S}^2_2$ under the homeomorphism; we define  ${O}=\{ ({p}^1_1,{ p}^1_2), \cdots, ({ p}^q_{1},{ p}^q_2) \}$, and with some abuse of notation we let $\Omega= \mathbb{S}^2_1 \cup \mathbb{S}^2_2$, and hence $O \subset \Omega \times \Omega$. For a single subject, we can model the streamline endpoints $O$ as a realization of an underlying point process on $\Omega \times \Omega$ with an unknown integrable intensity function $u: \Omega \times \Omega \mapsto [0,\infty)$ defined as follows. For any two measurable regions $E_1 \subset \Omega$ and $E_2 \subset \Omega$, denote $\#(E_1,E_2)$ as the counting process of the number of streamlines ending in $(E_1,E_2)$. Then $u$ satisfies
\begin{equation}\label{eqn:first_moment_assumption}
    \mathbb{E}\left[\#(E_1,E_2)\right]=\int_{E_1} \int_{E_2} u(\omega_1,\omega_2)d\omega_1d\omega_2 < \infty
\end{equation}
\par 
In this work, we are interested in the analysis of a random sample of the structural connectivity data from $N$ subjects: $O_1,...,O_N$. Modeling the replicated point processes with a single deterministic intensity function is likely insufficient to properly accommodate the population variability in connectivity patterns. Hence, it is reasonable to assume that the intensity function governing $O$ is itself a realization of an underlying random process, which we denote as $U$. That is, conditional on $U_i\sim U$, the first order moment of the point set $O_i$ satisfies Equation~\eqref{eqn:first_moment_assumption} with intensity function $U_i$. 
\par 
Denote $\mathcal{H}$ as the space of symmetric $L^2$ functions over $\Omega\times\Omega$, i.e., for $u \in \mathcal{H}$, $u(\omega_1,\omega_2) = u(\omega_2,\omega_1)$ and $\int_\Omega\int_\Omega u^2(\omega_1,\omega_2)d\omega_1d\omega_2<\infty$. We assume that $U$ has an associated measure $\mathbb{P}$ such that i) the mean function $\mathbb{E}_{\mathbb{P}}[U] = \mu\in \mathcal{H}$, ii) the process has finite second order moment $\mathbb{E}_{\mathbb{P}}[\|U\|_{L^{2}(\Omega\times\Omega)}^2]<\infty$, iii) the covariance function $C\left( (\omega_1,\omega_2), (\omega_1^{\prime},\omega_2^{\prime}) \right) := \text{Cov}\{U(\omega_1,\omega_2),U(\omega_1^{\prime},\omega_2^{\prime})\}$ is mean-square integrable and iv) $U$ is integrable almost surely. Under these conditions, $U$ is associated with a random density function, denoted $F_U(\omega_1, \omega_2) = U(\omega_1, \omega_2)/\int_{\Omega\times\Omega}U$. Define $Q$ as a distribution on $\mathbb{N}$ with finite moments modeling the total number of streamlines. $Q$ is related to the seeding procedure in fiber tracking \citep{girard2014towards,ambrosen2020validation}, \markrevised{e.g., with more seeds, more streamlines will be observed.  Conditional on $U_i$ and $q_i$, we assume that the $i$'th subject's observed streamline endpoints $O_i\overset{i.i.d.}{\sim} F_{U_{i}}$ and that $Q$ and $F_{U}$ are independent \citep{muller2013}.} 

\subsection{Subject-Level Estimate of Continuous Connectivity}\label{ssec:subj_level_estim}
Under the model described in Section~\ref{ssec:statistical_model}, conditional on $O_i$, we can form a \markrevised{pointwise estimate $\widehat{F_{U_{i}}}$ (and hence $\widehat{U}_i$) by local density estimation. In this work, we use the augmented symmetrized product heat kernel first proposed in \cite{moyer2017} for estimating the rate function of an inhomogeneous Poisson process model for a single continuous connectome. Several alternative pointwise smoothers have been proposed in the connectomics literature \citep{borovitskiy2021,mansour2022},} 
\markrevised{any of which could used instead while retaining compatibility with the remainder of our methodology. In the following, we provide an explicit formulation of chosen kernel smoother.} Let $H_{h}:\Omega\times\Omega\mapsto \mathbb{R}^{+} \cup \{0\}$ be the kernel function with bandwidth $h > 0$ defined by $H_h((\omega_1,\omega_2)|(p_1,p_2)) = \kappa_h(\omega_1, p_1)\kappa_h(\omega_2, p_2)$, where $\kappa_h$ is the spherical heat kernel \citep{chung2006} trivially extended to $\Omega$ by setting $\kappa_h(\omega_1,p_1) = 0$ if $\omega_1$ and $p_1$ are not on the same copy of $\mathbb{S}^2$. A point-wise estimate of $F_{U_i}$ and $U_i$ for the $i$-th subject is given by:
\begin{equation}\label{eqn:KDE_cc}
    \begin{aligned}
            \widehat{F_{U_{i}}}(\omega_1,\omega_2) &= \sum_{j=1}^{q_i} (H_{h}((\omega_1,\omega_2)|(p^{j}_{i1}, p^{j}_{i2}))+H_{h}((\omega_2,\omega_1)|(p^{j}_{i2} , p^{j}_{i1})))/2q_i, \\
            \widehat{U_i} &= q_i \widehat{F}_i(\omega_1,\omega_2).
    \end{aligned}
\end{equation}
\markrevised{In our real data analysis, $q_i$ is on the order of $10^6$, and so we can expect these point-wise estimates to be reasonable. Selection of the bandwidth $h$ can be performed using the cross-validation procedure described in \cite{moyer2017}.}

\section{Reduced-Rank Modeling of Continuous Connectivity}\label{sec:reduced_rank_model}
\markrevised{Given a set of continuous connectivity objects estimated using \eqref{eqn:KDE_cc}, this section proposes a set of novel methods for efficient joint representation and downstream statistical analyses. Proofs of all the results in this section are provided in Section~\ref{sec:theory_proofs} of the Supplemental Materials.}

\subsection{Functional Principal Components Analysis}\label{ssec:fpca}
Optimal reduced-rank representation for functional data has largely focused on the eigenfunctions of the covariance function $C$. Under the mean-square integrability assumption on $C$, Mercer's theorem guarantees the existence of a set of non-negative eigenvalues $\{\rho_k\}$ and associated orthonormal eigenfunctions $\{\psi_k\}$ such that 
$
C\left( (\omega_1,\omega_2), (\omega_1^{\prime},\omega_2^{\prime}) \right) = \overset{\infty}{\underset{k=1}{\sum}}\rho_k \psi_k(\omega_1,\omega_2)\psi_k(\omega_1^{\prime},\omega_2^{\prime}).
$
 By the Karhunen-Lo\`{e}ve theorem, the random function can be represented as
$
U(\omega_1, \omega_2) = \mu(\omega_1,\omega_2) + \sum_{k=1}^\infty Z_k\psi_k(\omega_1, \omega_2),
$ where $Z_k = \langle U - \mu, \psi_k\rangle_{L^2(\Omega\times\Omega)}$ are independent, mean zero random variables with $\mathbb{E}_{\mathbb{P}}[Z_k^2] = \rho_k$.  
The first $K$ $\psi_k$'s form an optimal (according to the mean integrated squared error) rank $K$ basis for representing realizations of $U$, making them appealing for forming parsimonious finite rank approximations. Of course, the eigenfunctions are unknown and must be estimated from the data, i.e., by functional principal components analysis (FPCA) \citep{ramsay2005}. A common approach to FPCA is to form a smoothed estimate $\widehat{C}$ and then perform a spectral decomposition over some finite-dimensional basis system or discretization of the domain \citep{silverman1996,yao2005}. However, this approach is infeasible in the current situation due to the curse of dimensionality, as $C$ is a symmetric positive semi-definite (PSD) function defined over an $8$-D manifold domain, i.e., $(\Omega \times \Omega) \times (\Omega \times \Omega)$. \markrevised{For instance, imposing a grid with $V = 4000$ vertices for $\Omega$, the resulting discretization of $C$ is a matrix of  $1.6*10^7 \times 1.6*10^7$ elements, the storage of which alone requires $\approx 8*10^{6} GB$, 
exceeding the computational capabilities of most currently existing computers.
} 
\par 
Approaches to FPCA for data on complicated non-Euclidean multidimensional domains are limited. \cite{lila2016} use finite elements to solve the best rank $K$ optimization problem, while \cite{muller2017} and \cite{lynch2018} assume a separable structure on $C$ to promote tractable estimation of the marginal eigenfunctions. The former still requires solving an optimization problem over $\mathcal{H}$, the space of symmetric functions on the product space of two 2-D submanifolds of $\mathbb{R}^3$, hence the curse of dimensionality is still problematic. The latter introduces assumptions that are hard to verify in practice, and can be inefficient when the true covariance is not separable \citep{consagra2021}. Instead, we propose an alternative data-driven procedure to create a reduced-rank function space, adapted to the distribution of $U$, that is both highly flexible and avoids the curse of dimensionality. 

\subsection{Data-Driven Reduced-Rank Function Spaces}\label{ssec:reduced_rank_model}
We begin with defining relevant notations. Without loss of generality, assume that $U$ has been centered, i.e., we consider the zero-mean process $\mathbb{E}_{\mathbb{P}}[U]=0$. Denote the set of unit $L^2(\Omega)$-norm functions, called the Hilbert sphere, $\mathbb{S}^{\infty}(\Omega) :=  \{\xi \in L^2(\Omega) : \left\| \xi \right\|_{L^{2}(\Omega)} = 1\}$.  Define the symmetric separable orthogonal function set of rank $K$ as
$$
\mathcal{V}_{K} = \{\xi_k\otimes\xi_k: \xi_k \in \mathbb{S}^{\infty}(\Omega), \langle \xi_k, \xi_j\rangle_{L^{2}(\Omega)} = \delta_{kj},\text{ for }k = 1,2,...,K\},
$$
where $\xi\otimes\xi(\omega_1,\omega_2) := \xi(\omega_1)\xi(\omega_2)$ and $\delta_{ik}$ is the Kronecker delta. 
\par 
For any $K$, there are an infinite number of such sets. We propose to construct a $\mathcal{V}_{K}$ adapted to the distribution of $U$ utilizing a greedy learning procedure. Specifically, given a sample of $N$ independent realizations of $U_i \sim U$, we iteratively construct $\mathcal{V}_K$ by repeating the following steps: 
\begin{equation}\label{eqn:empirical_greedy_algorithm}
\begin{aligned}
        \xi_k &= \sup_{\xi\in\mathbb{S}^{\infty}(\Omega)}N^{-1}\sum_{i=1}^N\left[|\left\langle R_{k-1,i}, \xi\otimes\xi\right\rangle_{L^2(\Omega\times\Omega)}|^2\right] \\
        \quad \mathcal{V}_k &= \mathcal{V}_{k-1} \cup \{\xi_k \otimes \xi_k\}\\
        \quad R_{k,i} &= U_i - P_{\mathcal{V}_k}(U_i), \\ 
\end{aligned}
\end{equation}
for $k=1,...,K$, where $P_{\mathcal{V}_k}$ is the orthogonal projection operator onto $\text{span}\left(\mathcal{V}_k\right)$ and the process is initialized with $R_{0,i} = U_i$ and $\mathcal{V}_0=\emptyset$.
\par 
We would like to characterize the theoretical performance of the resulting approximation space, $\text{span}\left(\mathcal{V}_k\right)$, for representing the continuous connectivity $U_i$. \markrevised{The following theorem, which was initially presented in \cite{Consagra2022},  bounds the asymptotic mean $L^2$ representation error of the low-rank space constructed from the greedy updates in \eqref{eqn:empirical_greedy_algorithm} as a function of $K$.}
\begin{theorem} [\cite{Consagra2022}] \label{thrm:sample_convergence_rate}
Let $U_1, ..., U_N$ be i.i.d. samples from $U$. Under Assumptions~\ref{assm:smoothness_class}, \ref{assm:moment_bound} and \ref{assm:decay_rate} \markrevised{in the Supplemental Materials}, with probability 1:
$$
N^{-1}\sum_{i=1}^N \left\|R_{K,i}\right\|_{L^2(\Omega\times\Omega)}^2 \le   \frac{B_1^2\left(\sum_{k=1}^\infty\sqrt{\rho_k}\right)^2}{K+1}\quad\text{as}\quad N\rightarrow\infty,
$$
\markrevised{where constant $B_1<\infty$ is related to the tail behavior of $C$.}
\end{theorem}
In general, $\xi_k\otimes\xi_k\neq\psi_k$, so $\text{span}\left(\mathcal{V}_K\right)$ is not the optimal rank $K$ space for representing realizations of $U$. 
However, $\mathcal{V}_K$ offers a desirable trade-off between flexibility and computational complexity. In regard to the latter, notice that \eqref{eqn:empirical_greedy_algorithm} requires solving an optimization problem over $L^2(\Omega)$, while estimating the eigenfunctions requires an optimization problem over $\mathcal{H}\subset L^{2}(\Omega\times\Omega)$. 
This simplification is critical in practice, as optimization for functions directly in $L^{2}(\Omega\times\Omega)$ squares the number of unknown parameters that need to be estimated, resulting in enormous computational difficulties due to the high dimensionality of the marginal space $\Omega$. The flexibility of $\mathcal{V}_K$ is demonstrated in Theorem~\ref{thrm:sample_convergence_rate}, which establishes both asymptotic completeness and error bounds as a function of the rank. 

\subsection{Statistical Analysis of Reduced-Rank Continuous Connectivity}\label{ssec:inference}
We can approximate any function in $\mathcal{H}$ as a linear combination of basis functions in $\mathcal{V}_K$, thereby mapping the infinite dimensional continuous connectivity $U_i$ to a low-dimensional vector consisting of the coefficients of the basis expansion. As such, any continuous connectivity $U_i \in \mathcal{H}$ can be identified with a $K$-dimensional Euclidean vector $\boldsymbol{s}_i := [S_{i1},...,S_{iK}]^T$, with $P_{\mathcal{V}_K}(U_i) = \sum_{k=1}^K S_{ik}\xi_k\otimes\xi_k$. Owing to the orthonormality of the basis functions in $\mathcal{V}_K$, the mapping $U_i\mapsto \boldsymbol{s}_i$ is an isometry between $\left(\text{span}(\mathcal{V}_K), \langle\cdot,\cdot\rangle_{L^{2}(\Omega\times\Omega)}\right)$ and $\left(\mathbb{R}^K, \langle\cdot,\cdot\rangle_2\right)$, where $\langle\cdot,\cdot\rangle_2$ is the standard Euclidean metric. Therefore, we can properly embed the continuous connectivity into a $K$-dimensional Euclidean vector space and utilize a multitude of existing tools from multivariate statistics for analysis of $U_i$. For the remainder of this section, we discuss how to perform several canonical tasks of interest in computational neuroscience under our continuous connectivity data representation. 
\par

\noindent{{\bf \textit{Trait Prediction}}: It is often desirable to use the structural connectivity to predict some trait of interest. This can be accommodated under our framework simply by using the embedding vectors $\boldsymbol{s}_i$ as features in an appropriate predictive model.}
\par
\noindent{{\bf \textit{Global Hypothesis Testing}}: Another important application is the assessment of global differences in structural connectivity between two groups. Given two samples of continuous connectivity: $\{U_{11}, ..., U_{1N_{1}}\},$ and $\{U_{21}, ..., U_{2N_{2}}\}$, we are interested in testing $U_1 \overset{\text{dist}}{=} U_2$, where $U_{li}\sim U_l$, and $l=1,2$ denotes the group membership. Using the continuous connectivity embedding, this testing problem is translated into a two-group test on the coefficients $\mathcal{S}_1:=\{\boldsymbol{s}_{11}, ..., \boldsymbol{s}_{1N_{1}}\}$ v.s. $\mathcal{S}_2:=\{\boldsymbol{s}_{21},...,\boldsymbol{s}_{2N_{2}}\}$. A non-parametric test to assess whether $\mathcal{S}_1$ and $\mathcal{S}_2$ are independent samples from the same distribution can be performed using the Maximum Mean
Discrepancy (MMD) test statistic \citep{gretton2012}.}
\par
\noindent{{\bf \textit{Identifying Local Subnetwork Differences Between Groups}}: In addition to testing global connectivity difference, it is often of interest to identify where this difference manifests in the brain. In atlas-based approaches, this problem corresponds to the identification of brain subnetworks related to phenotypic traits of interest, which is a significant area of interest in modern structural connectome analysis \citep{chung2021}. Analogously, under the continuous connectivity framework, we want to identify regions of the domain $\Omega\times\Omega$ where the continuous connectivity is different between groups. This is known as a \textit{local inference} problem in the FDA literature.  One common approach to local inference is to represent the functional data using some basis system with local support, e.g., splines or wavelets, and then test differences on the coefficients of the expansion \citep{pini2016}. The local support property of the basis is critical as it allows the effects of the coefficients to be localized on the domain. To perform local inference for continuous connectivity, we need to construct locally supported $\xi_k$. Therefore, we augment our basis learning procedure to include an optional constraint to promote basis with sparse support (more details are presented in Section~\ref{sec:methods}).}
\par 
Let $U_l\approx P_{\mathcal{V}_{K}}(U_{l}) = \sum_{k=1}^K S_{lk}\xi_k\otimes\xi_k$, with $l=1,2$ again being the group label. The following theorem illustrates how we can link hypothesis tests on coefficient differences to the identification of subnetworks that differ between groups.
\begin{theorem}\label{thrm:continuous_subnetwork_discovery}
Define a collection of $K$ testing problems on the coefficients
\begin{equation}\label{eqn:testing_problem}
    H_{k}^0: S_{1k} \overset{\text{dist}}{=} S_{2k} \qquad H_{k}^a: S_{1k} \overset{\text{dist}}{\neq} S_{2k}, \qquad \text{for } k = 1,...,K.
\end{equation}
Denote the index set $I=\{k\in\{1,...,K\}:H_{k}^a\text{ is true}\}$, and define
$$
\text{Supp}(\xi_k) := \{ \omega_1 \in \Omega:\xi_k(\omega_1) \neq 0\},\quad\text{Supp}(\xi_k\otimes\xi_k) := \{ (\omega_1,\omega_2) \in \Omega\times\Omega:\xi_k(\omega_1)\xi_k(\omega_2) \neq 0\}
$$
and $\mathcal{C} := \bigcup_{k\in I}\text{Supp}\left(\xi_k\otimes\xi_k\right) \subset\Omega\times\Omega$.
A non-empty $I$ implies the following point-wise condition: 
$
\exists (\omega_1,\omega_2) \in \mathcal{C} \text{ such that } U_1(\omega_1,\omega_2) \overset{\text{dist}}{\neq} U_2(\omega_1,\omega_2).
$
\end{theorem}
The set $\mathcal{C}$ covers brain regions where some $U_1(\omega_1,\omega_2) \overset{\text{dist}}{\neq} U_2(\omega_1,\omega_2)$.
Since the group labels are exchangeable under $H_{k}^{0}$, the set of $K$ tests in Equation~\eqref{eqn:testing_problem} used to construct $\mathcal{C}$ can be performed using simple permutations and the resulting p-values corrected using \cite{holm1979} to control the family-wise error rate (FWER) \markrevised{ at a pre-specified level $\alpha\in[0,1]$. Denote $\mathcal{C}_{\alpha}$ to be the subnetwork cover constructed using the proposed testing scheme and define the false coverage probability (FCP) as follows:
\begin{equation}\label{eqn:FCP}
    \text{FCP}(\mathcal{C}_{\alpha}) := \mathbb{P}\left[\mathcal{C}_{\alpha}\neq \emptyset\text{ and } U_1(\omega_1, \omega_2) \overset{\text{dist}}{=} U_2(\omega_1, \omega_2), \forall (\omega_1,\omega_2)\in \mathcal{C}_{\alpha}\right].
\end{equation}
The FCP quantifies the probability that a non-empty $\mathcal{C}_{\alpha}$ is formed, but it does not cover any significantly different edges. The following theorem characterizes the control of the FCP under the proposed testing procedure:
\begin{theorem}\label{thm:FCP_control}
The proposed testing procedure controls the false coverage proportion at level $\alpha$, that is, FCP$(\mathcal{C}_{\alpha})\le \alpha$.
\end{theorem} 
}


\section{Estimation and Implementation Details}\label{sec:methods}
\markrevised{In this section, we derive an alternating optimization scheme to estimate the reduced-rank function space in \eqref{eqn:empirical_greedy_algorithm} from the observed data  
and discuss associated implementation details and hyperparamter selection.}
\subsection{Deriving the Optimization Problem}
Using the orthogonal constraint for elements in $\mathcal{V}_K$, the $k$-th greedy search of $\xi_k$ in \eqref{eqn:empirical_greedy_algorithm} can be equivalently formulated as:
\begin{equation}\label{eqn:empirical_optm_reformulated}
\begin{aligned}
    \hat{\xi}_k = &\sup_{\xi \in L^2(\Omega)}\frac{1}{N}\sum_{i=1}^N\left\langle R_{k-1,i}, \xi\otimes\xi\right\rangle_{L^2(\Omega\times\Omega)}^2 \\
    \textrm{s.t.} \quad & \langle \xi, \xi_j \rangle_{L^{2}(\Omega)} = 0 \text{ for }j=1,2,...,k-1, \quad \left\|\xi\right\|_{L^2(\Omega)} = 1 \\
\end{aligned}
\end{equation}
Unfortunately, problem \eqref{eqn:empirical_optm_reformulated} is intractable due to the infinite-dimensional search space $L^{2}(\Omega)$. In what follows, we derive a discrete approximation of problem \eqref{eqn:empirical_optm_reformulated} that facilities efficient computation by the algorithm proposed in Section~\ref{ssec:algorithm}.
\par 
For tractable computation of the $L^2(\Omega\times\Omega)$ inner product, we impose a dense mesh grid on $\Omega\times\Omega$. Let $\boldsymbol{X} = (\boldsymbol{x}_{11}, ..., \boldsymbol{x}_{1n_{1}}, \boldsymbol{x}_{22}, ..., \boldsymbol{x}_{2n_{2}})^\intercal \in \mathbb{R}^{(n_1+n_2)\times 3}$ be the vertices on the marginal grid. Note that since $\Omega$ is a union of two 2-spheres, $\{ \boldsymbol{x}_{11}, ..., \boldsymbol{x}_{1n_{1}}\}$ are points on $\mathbb{S}_1^2$ and $\{\boldsymbol{x}_{22}, ..., \boldsymbol{x}_{2n_{2}}\}$ are points on $\mathbb{S}_2^2$.  For subject $i$, denote the high-resolution symmetric continuous connectivity matrix:
$\boldsymbol{Y}_{i} =\begin{bmatrix}
                            \boldsymbol{Y}_{i,11} & \boldsymbol{Y}_{i,12}  \\
                            \boldsymbol{Y}_{i,12}^{\intercal}& \boldsymbol{Y}_{i,22}
                        \end{bmatrix}$,
where $\boldsymbol{Y}_{i,d_{1}d_{2}} \in \mathbb{R}^{n_{d_{1}}\times n_{d_{2}}}$ contains elements $\widehat{U}_i(\boldsymbol{x}_{d_{1}\cdot}, \boldsymbol{x}_{d_{2}\cdot})$. The inner product in \eqref{eqn:empirical_optm_reformulated} can now be approximated as 
\begin{equation}\label{eqn:discrete_l2_norm}
    \left\langle \widehat{R}_{k,i}, \xi\otimes\xi\right\rangle_{L^{2}(\Omega\times\Omega)}^2 \approx (n_1n_2)^{-1}\left\langle \boldsymbol{R}_{k,i}, \boldsymbol{\xi}\otimes\boldsymbol{\xi}\right\rangle_{F}^2,
\end{equation}
where $\langle,\rangle_{F}$ is the Frobenius inner product, $\boldsymbol{\xi}_j\in\mathbb{R}^{n_1+n_2}$ are the vectors of evaluations of $\xi_j$ on $\boldsymbol{X}$, $\boldsymbol{R}_{k,i} = \boldsymbol{Y}_i - \bar{\boldsymbol{Y}}  - \sum_{j=1}^k s_{ij}\boldsymbol{\xi}_j\otimes\boldsymbol{\xi}_j$, with 
\begin{equation}\label{eqn:coef_rewrite}
    s_{ij} = (n_1n_2)^{-1}\left\langle\boldsymbol{R}_{j-1,i},\boldsymbol{\xi}_j\otimes\boldsymbol{\xi}_j\right\rangle_{F}, 
\end{equation}
and $\bar{\boldsymbol{Y}}$ is the mean of the $\boldsymbol{Y}_i$'s, \markrevised{see Supplemental Section~\ref{ssec:algoImp} for more details.}

We now handle the infinite dimensional search space of $\xi_k\in L^2(\Omega)$ through basis expansion. As $\Omega$ is a union of two 2-spheres, we can trivially extend any complete basis system for $L^2(\mathbb{S}^2)$, denoted $\{\phi_{j}\}_{j=1}^\infty$, to $L^2(\Omega)$, since $L^2(\Omega) = L^2(\mathbb{S}^2)\cup L^2(\mathbb{S}^2) = \text{span}(\{\phi_{j}\}_{j=1}^\infty)\cup \text{span}(\{\phi_{j}\}_{j=1}^\infty)$.  Denoting the $M_d$-dimensional truncation $\boldsymbol{\phi}_{M_{d}} = (\phi_1, ..., \phi_{M_{d}})^\intercal$ (where $d \in \{1,2\}$ indexes the two spheres in $\Omega$), $\xi_k$ is approximated as 
\begin{equation}\label{eqn:basis_expansion}
 \xi_{k}(\omega) = \boldsymbol{c}_{1,k}^\intercal\boldsymbol{\phi}_{M_{1}}(\omega)\mathbb{I}\{\omega\in\mathbb{S}^2_1\} + \boldsymbol{c}_{2,k}^\intercal\boldsymbol{\phi}_{M_{2}}(\omega)\mathbb{I}\{\omega\in\mathbb{S}^2_2\},
\end{equation}
where the $\boldsymbol{c}_{d,k} \in \mathbb{R}^{M_{d}}$ are the vectors of coefficients with respect to the basis $\boldsymbol{\phi}_{M_{d}}$. Hence, an element in $L^2(\Omega)$ can be represented as $\boldsymbol{c}_k = (\boldsymbol{c}_{1,k}^\intercal,\boldsymbol{c}_{2,k}^\intercal)^\intercal$ with basis functions $\boldsymbol{\phi}_M = (\boldsymbol{\phi}_{M_{1}}^\intercal,\boldsymbol{\phi}_{M_{2}}^\intercal)^\intercal$, where $M = M_{1}+M_{2}$. In theory, $\{\phi_{j}\}_{j=1}^\infty$ can be taken as any complete basis system for $L^2(\mathbb{S}^2)$. In practice, we use spherical splines of degree 1, due to their appealing properties \markrevised{(see Supplemental Materials Section~\ref{sec:marg_splines} for more discussion.)}
\par 
In applications, it may be desirable to incorporate some constraints on the learned basis functions $\{\xi_k\}$. For example, we want to promote smoothness in the $\xi_k$ to ameliorate the effect of discretization and/or estimate locally supported $\xi_k$ to facilitate the local inference procedure discussed in Section~\ref{ssec:inference}. Smoothness can be achieved through a regularization term that penalizes the candidate solution's ``roughness'', measured using the integrated quadratic variation: $\text{Pen}_{QV}(\xi_k) = \int_{\Omega}\|\nabla_{\Omega}\xi_k(\omega)\|^2d\omega$. Owing to the local support of the spherical spline basis functions, locally supported $\xi_k$ can be achieved by encouraging sparsity in the $\boldsymbol{c}_k$'s. In particular, our estimation procedure incorporates an optional constraint set: $\{\boldsymbol{c}\in\mathbb{R}^{M}:\|\boldsymbol{c}\|_0\le n_{\alpha_{2}}\}$, where $n_{\alpha_{2}}$ is a tuning parameter controlling the level of localization of the basis functions. 
\par 
Define $\boldsymbol{\Phi}_d\in\mathbb{R}^{n_d \times M_{d}}$ to be the matrix of evaluations of $\boldsymbol{\phi}_{M_{d}}$ over $\{ \boldsymbol{x}_{d1}, ..., \boldsymbol{x}_{dn_{1}}\}$ and denote the matrix of inner products $\boldsymbol{J}_{d} = \int_{\mathbb{S}^2}\boldsymbol{\phi}_{M_{d}}\boldsymbol{\phi}_{M_{d}}^\intercal$. Define the block matrices 
$$
\boldsymbol{\Phi} = \begin{bmatrix} 
                       \boldsymbol{\Phi}_1 & \boldsymbol{0} \\
                        \boldsymbol{0} &\boldsymbol{\Phi}_2
                \end{bmatrix}, \quad \boldsymbol{J}_{\phi} = \begin{bmatrix} 
\boldsymbol{J}_{1} & \boldsymbol{0} \\
\boldsymbol{0} &  \boldsymbol{J}_{2}
\end{bmatrix}.
$$
After incorporating the smoothness and local support constraint, the final formulation of \eqref{eqn:empirical_optm_reformulated}
is given by 
\begin{equation}\label{eqn:empirical_optm_reformulated_discretized}
\begin{aligned}
    \hat{\boldsymbol{c}}_k = &\underset{\boldsymbol{c} \in \mathbb{R}^{M}}{\text{ argmax }}\frac{1}{Nn_1n_2}\sum_{i=1}^N\left\langle \boldsymbol{R}_{k-1,i}, (\boldsymbol{\Phi}\boldsymbol{c})\otimes (\boldsymbol{\Phi}\boldsymbol{c})\right\rangle_{F}^2 -  \alpha_1 \boldsymbol{c}^\intercal\boldsymbol{Q}_{\boldsymbol{\phi}}\boldsymbol{c} \\
    \textrm{s.t.} \quad & \boldsymbol{c}^\intercal\boldsymbol{J}_{\phi}\boldsymbol{c} = 1, \quad \boldsymbol{c}^\intercal\boldsymbol{J}_{\phi}\boldsymbol{c}_j  = 0 \text{ for }j=1,2,...,k-1, \quad \left\|\boldsymbol{c}\right\|_{0} \le n_{\alpha_{2}} \\
\end{aligned}
\end{equation}
where $\alpha_1, n_{\alpha_{2}} > 0$ are hyperparameters used to control the smoothness and local support size, respectively, and $ \boldsymbol{c}^\intercal\boldsymbol{Q}_{\boldsymbol{\phi}}\boldsymbol{c}$ encodes $\text{Pen}_{QV}(\xi_k)$ after basis expansion (see the Supplemental Materials Section~\ref{ssec:pen_matrix_construction} for more details).  Efficient algorithms exist for evaluating spherical splines, computing their directional derivatives as well as performing integration \citep{schumaker2015}, hence the matrices $\boldsymbol{J}_{\boldsymbol{\phi}}$ and $\boldsymbol{Q}_{\boldsymbol{\phi}}$ can be constructed cheaply.

\subsection{Algorithm}\label{ssec:algorithm} 
\begin{algorithm}[!ht]
\caption{AO Algorithm to approximate \eqref{eqn:empirical_optm_reformulated_discretized_AO}}\label{alg:rank_1_ssb}
\begin{algorithmic}[1]
\State \textbf{Input}: $\text{Tensor }\mathcal{Y}; \text{ matrices }\boldsymbol{J}_{\boldsymbol{\phi}}, \boldsymbol{Q}_{\boldsymbol{\phi}}, \boldsymbol{U},\boldsymbol{V},\boldsymbol{D};  \text{ rank }K; \text{ hyperparameters } \alpha_1, n_{\alpha_{2}}$
\State \textbf{Output}: $\boldsymbol{C}, \boldsymbol{S}$
\State \textbf{Initialize}: $\mathcal{G}_{0} \gets  \mathcal{G} := \mathcal{Y}\times_1\boldsymbol{U}^\intercal\times_2\boldsymbol{U}^\intercal$; $\boldsymbol{C} \gets \emptyset$; $\boldsymbol{S}\gets \emptyset$; $\boldsymbol{P}_0\gets \boldsymbol{0}_{M\times M}$
\For{$k=1,...K$}
    \State \textbf{Initialize} $\boldsymbol{s}_k^{(0)}$  using the leading left singular vector of the mode-3 matricization of $\mathcal{G}_{k-1}$
    \State \textbf{Initialize} $\boldsymbol{c}_k^{(0)}$ using  $\boldsymbol{V}\boldsymbol{D}^{-1}$ product with the leading left singular vector of the mode-1 matricization of $\mathcal{G}_{k-1}$
    \While{\text{Not Converged}}
        \State 
        \State \textbf{Update} $\boldsymbol{c}_k^{(t+1)}$ according to \eqref{eqn:ctilde_update} using initial guess $\boldsymbol{c}_k^{(t)}$
        \State \textbf{Update} $\boldsymbol{s}_k^{(t+1)}$ according to \eqref{eqn:s_update}
    \EndWhile
    \State Compute $\boldsymbol{P}_k$ using \eqref{eqn:update_P}
    \State $\mathcal{G}_{k} \gets \mathcal{G}_{k-1}- \left(\boldsymbol{D}\boldsymbol{V}^\intercal\boldsymbol{c}_k^{(t+1)}\right) \otimes \left(\boldsymbol{D}\boldsymbol{V}^\intercal\boldsymbol{c}_k^{(t+1)}\right) \otimes  \boldsymbol{s}_k^{(t+1)}$
    \State $\boldsymbol{C} \gets \left[\boldsymbol{C};\boldsymbol{c}_k^{(t+1)}\right]$
    \State $\boldsymbol{S} \gets \left[\boldsymbol{S}; \boldsymbol{s}_k^{(t+1)}\right]$
\EndFor

\end{algorithmic}
\end{algorithm}

Denote $\mathcal{Y}$ as the $(n_1 + n_2) \times (n_1 + n_2) \times N$-dimensional semi-symmetric tensor obtained from a mode-3 stacking of the $\boldsymbol{Y}_i-\bar{\boldsymbol{Y}}$. 
The computational grid $\boldsymbol{X}$ can be made arbitrarily dense, rendering $\mathcal{Y}$ an enormously high dimensional ``functional tensor'' object. This prohibits a standard tensor decomposition applied to $\mathcal{Y}$ and necessitates the development of our efficient algorithm discussed below. 
\par 
The optimization problem \eqref{eqn:empirical_optm_reformulated_discretized} is a constrained maximization of a degree 4 polynomial with $M \ge 3$ variables and thus is NP-hard \citep{hou2014}. To derive a computationally tractable algorithm, we first notice that by combing Equations~\eqref{eqn:coef_rewrite} and \eqref{eqn:basis_expansion}, we have that 
\begin{equation}\label{eqn:coef_rewrite_basis}
    s_{ik} \approx (n_1n_2)^{-1}\langle \boldsymbol{R}_{k-1,i}, \left(\boldsymbol{\Phi}\boldsymbol{c}_k\right)\otimes\left(\boldsymbol{\Phi}\boldsymbol{c}_k\right)\rangle_{F}.  \\
\end{equation}
Using approximation \eqref{eqn:coef_rewrite_basis} and introducing the associated auxiliary variable $\boldsymbol{s}=(s_1,...,s_{N})^\intercal\in\mathbb{R}^{N}$, \eqref{eqn:empirical_optm_reformulated_discretized} can be equivalently defined as 
\begin{equation}\label{eqn:empirical_optm_reformulated_discretized_AO}
\begin{aligned}
    \hat{\boldsymbol{c}}_k = &\underset{\boldsymbol{c} \in \mathbb{R}^{M}}{\text{ argmax }}\sum_{i=1}^Ns_i\left\langle\boldsymbol{R}_{k-1,i}, \left(\boldsymbol{\Phi}\boldsymbol{c}\right)\otimes\left(\boldsymbol{\Phi}\boldsymbol{c}\right)\right\rangle_{F} -  \alpha_1 \boldsymbol{c}^\intercal\boldsymbol{Q}_{\boldsymbol{\phi}}\boldsymbol{c} \\
    \textrm{s.t.} \quad & \boldsymbol{c}^\intercal\boldsymbol{J}_{\phi}\boldsymbol{c} = 1, \boldsymbol{c}^\intercal\boldsymbol{J}_{\phi}\boldsymbol{c}_j  = 0 \text{ for }j=1,2,...,k-1,\\
    \quad & \left\|\boldsymbol{c}\right\|_{0} \le n_{\alpha_{2}}  , \quad s_i = \langle \boldsymbol{R}_{k-1,i}, \left(\boldsymbol{\Phi}\boldsymbol{c}\right)\otimes\left(\boldsymbol{\Phi}\boldsymbol{c}\right)\rangle_{F}, \text{ for } i=1,...,N.
\end{aligned}
\end{equation}
\par 
We apply an alternating optimization (AO) scheme to \eqref{eqn:empirical_optm_reformulated_discretized_AO}, iteratively maximizing $\boldsymbol{c}$ given $\boldsymbol{s}$ and vice versa. A data-reduction transformation based on the singular value decomposition $\boldsymbol{\Phi}_d = \boldsymbol{U}_d\boldsymbol{D}_d\boldsymbol{V}_d^\intercal$ is utilized in order to avoid computations that scale with the number of grid points (the dimension of $\boldsymbol{Y}$). Define the block matrices 
$$
\boldsymbol{V} = \begin{bmatrix} 
                        \boldsymbol{V}_1 & \boldsymbol{0} \\
                        \boldsymbol{0} & \boldsymbol{V}_2
                \end{bmatrix}, \qquad 
                 \boldsymbol{D} = \begin{bmatrix} 
                        \boldsymbol{D}_1 & \boldsymbol{0} \\
                        \boldsymbol{0} & \boldsymbol{D}_2
                \end{bmatrix}, \qquad \boldsymbol{U} = \begin{bmatrix} 
                        \boldsymbol{U}_1 & \boldsymbol{0} \\
                        \boldsymbol{0} & \boldsymbol{U}_2
                \end{bmatrix},
$$
and define $\boldsymbol{G}_{k,i} := \boldsymbol{U}^\intercal\boldsymbol{R}_{k,i}\boldsymbol{U}\in\mathbb{R}^{M\times M}$, which is the transformed residual of $\boldsymbol{R}_{k,i}$ for the $i$'th subject. 
By the properties of the Frobenius inner product, we have $\left\langle\boldsymbol{R}_{k-1,i}, \boldsymbol{\Phi}\boldsymbol{c}\otimes\boldsymbol{\Phi}\boldsymbol{c}\right\rangle_{F}=\left\langle\boldsymbol{G}_{k-1,i}, \boldsymbol{D}\boldsymbol{V}^\intercal\boldsymbol{c}\otimes\boldsymbol{D}\boldsymbol{V}^\intercal\boldsymbol{c}\right\rangle_{F}$. Crucially, this identity can be used in problem \eqref{eqn:empirical_optm_reformulated_discretized_AO} to avoid the computational burden of storing and computing with matrices of dimension $(n_1+n_2)\times(n_1+n_2)$.
\par 
Define the $M\times M \times N$-dimensional tensor $\mathcal{G}_{k}$ to be the mode-3 stacking of the $\boldsymbol{G}_{k,i}$. Then, by definition of the $d$-mode tensor-matrix multiplication (denoted $\times_{d}$), $\mathcal{G}_{0} := \mathcal{Y}\times_1\boldsymbol{U}^\intercal\times_2\boldsymbol{U}^\intercal$. Under the AO scheme, the update for the block variable $\boldsymbol{c}_k$ at the $t+1$ iteration is given by 
\begin{equation}\label{eqn:ctilde_update}
\begin{aligned}
    \boldsymbol{c}^{(t+1)} = \underset{\boldsymbol{c}\in\mathbb{R}^{M}}{\text{max}} &\quad \boldsymbol{c}^\intercal\left[\boldsymbol{V}\boldsymbol{D}\left(\boldsymbol{I} - \boldsymbol{P}_{k-1}\right)\left[\mathcal{G}_{k-1}\times_3\boldsymbol{s}^{(t)} - \alpha_1\boldsymbol{D}^{-1}\boldsymbol{V}^\intercal\boldsymbol{Q}_{\boldsymbol{\phi}}\boldsymbol{V}\boldsymbol{D}^{-1}\right]\left(\boldsymbol{I} - \boldsymbol{P}_{k-1}^{\intercal}\right)\boldsymbol{D}\boldsymbol{V}^\intercal\right]\boldsymbol{c} \\
    & \textrm{s.t.} \quad \boldsymbol{c}^\intercal\boldsymbol{J}_{\boldsymbol{\phi}}\boldsymbol{c} = 1, \quad \left\|\boldsymbol{c}\right\|_{0} \le n_{\alpha_{2}} , \\
\end{aligned}
\end{equation}
where 
\begin{equation}\label{eqn:update_P}
        \boldsymbol{P}_{k-1} = \boldsymbol{D}\boldsymbol{V}^\intercal\boldsymbol{C}_{k-1}\left[\boldsymbol{C}_{k-1}^{\intercal}\boldsymbol{J}_{\boldsymbol{\phi}}\boldsymbol{C}_{k-1}\right]^{-1}\boldsymbol{C}_{k-1}^{\intercal}\boldsymbol{J}_{\boldsymbol{\phi}}\boldsymbol{V}\boldsymbol{D}^{-1},  \quad \boldsymbol{C}_{k-1} := \left[\boldsymbol{c}_1, ..., \boldsymbol{c}_{k-1}\right],
\end{equation}
and the $\boldsymbol{c}_{j}$'s are from the previous $k-1$ selections. For $\boldsymbol{s}$, the update at the $t+1$ iteration is given in closed form by
\begin{equation}\label{eqn:s_update}
    \boldsymbol{s}^{(t+1)} = \mathcal{G}_{k-1} \times_1 (\boldsymbol{D}\boldsymbol{V}^\intercal\boldsymbol{c}_k^{(t+1)})\times_2 (\boldsymbol{D}\boldsymbol{V}^\intercal\boldsymbol{c}_k^{(t+1)}).
\end{equation}
For a detailed derivation of \eqref{eqn:ctilde_update} and \eqref{eqn:s_update}, see Section~\ref{ssec:algoImp} of the Supplementary Materials.
\par 
If no sparsity constraint is employed, \eqref{eqn:ctilde_update} reduces to finding the leading generalized eigenvector of a symmetric matrix. Since $\boldsymbol{J}_{\boldsymbol{\phi}}$ is symmetric and positive definite, this problem has a unique solution. Since \eqref{eqn:s_update} is trivially unique, the AO scheme of Algorithm~\ref{alg:rank_1_ssb} is guaranteed to converge to a stationary point of the objective by the existence and uniqueness property \citep{bezdek2003}.
\par 
Alternatively, with the sparsity constraint active, \eqref{eqn:ctilde_update} is recognized as a sparse generalized eigenvalue problem (SGEP), for which several recent methods could be applied to obtain a solution 
\citep{yuan2011,tan2018,jung2019,cai2021}.
\markrevised{To establish existence and uniqueness of a SGEP, additional assumptions on the problem structure are required \citep{cai2021}, and thus we cannot typically guarantee the convergence of the resulting AO scheme. In experiments not reported, we found such convergence issues to be problematic in the very sparse (small $n_{\alpha_{2}}$) regime, which corresponds to learning basis functions to detect highly localized effects. Therefore, in the interest of both computational speed and algorthmic stability, we employ an alternative two-stage procedure, in which we first obtain the non-sparsity constrained rank-1 updated $\boldsymbol{c}_{k}$ and then keep only the $n_{\alpha_{2}}$ largest elements in absolute magnitude. This ``single iteration thresholding'' corresponds to running a single iteration of the truncated power method from \cite{yuan2011} to sparsify the final estimate. Though somewhat ad-hoc, this method was found to be faster and more stable than iterative solving of the SGEP, while also allowing a fast and simple way to select $n_{\alpha_{2}}$ (discussed in Section~\ref{ssec:hyper_param_select}). Automatic hyperparameter selection for iterative sparse reconstructions often employ an adaptive BIC-like criteria \citep{allen2012} which, while theoretically satisfying, we found to further exacerbate issues when embedded into the AO-algorithm updates.}
\par 
Algorithm~\ref{alg:rank_1_ssb} provides pseudocode for the proposed AO scheme. Notice that the only computation that scales with the number of grid points is the transformation of the functional tensor $\mathcal{Y}$ into the tensor $\mathcal{G}$, which happens only once during the initialization of the procedure. Since typically $M \ll n_1+n_2$, this results in an enormous savings in computation when compared to an alternative approach of performing tensor decomposition directly on $\mathcal{Y}$.
Initializations for $\boldsymbol{c}^{(0)}$ and $\boldsymbol{s}^{(0)}$  are obtained using the leading left singular vectors of the mode-1 and mode-3 matricization of $\mathcal{G}_{k-1}$, see Supplementary Materials Section~\ref{ssec:algoImp} for justification.

\subsection{Hyperparameter Selection}\label{ssec:hyper_param_select}

\noindent{{\bf \textit{Smoothness and Sparsity}}: Algorithm~\ref{alg:rank_1_ssb} requires specifying the penalty parameters $\alpha_1$ (dictating the degree of smoothness of the solution), and $n_{\alpha_{2}}$ (determining the size of the support of $\xi_k$). In practice, $\alpha_1$ can be selected through cross-validation on the rank-1 approximation. In our real data analysis, we found small values were preferable, likely due to the fact that the functional data has already been smoothed via KDE over a dense grid. The local support parameter $ n_{\alpha_{2}}$ can be used if locally supported basis functions are desirable, e.g., for identifying local subnetworks that are different between groups, and can be tuned according to the desired sparsity level. \markrevised{If automatic sparsity parameter selection is desired, notice that we may equivalently formulate the problem as finding a numerical threshold $\tau_{k} > 0$ below which $|\boldsymbol{c}_k| \le  \tau_{k}$ is set to zero. To determine $\tau_{k}$, we propose to first apply univariate convex clustering to the elements of $\boldsymbol{c}_k$, select the cluster with centroid closest to 0, and set $\tau_{k}$ to be the absolute maximum over the zero-cluster elements. Similar clustering-based approaches have been used to remove estimates of small non-zero values in the high-dimensional sparse regression literature \citep{qiu2019}.}
\par 
\noindent{{\bf \textit{Rank Selection}}: It is also important to select both the marginal ranks $M_1,M_2$
and the global rank $K$. The global rank can be selected using a threshold criterion on the proportion of variance explained, estimated using the transformed tensor by $\left\|\mathcal{G}_{K}\right\|_{F}^2/\left\|\mathcal{G}\right\|_{F}^2$. Similarly, selecting the marginal ranks can be accomplished using a threshold criterion on $\|\mathcal{Y}\times_1\boldsymbol{U}^\intercal\times_2\boldsymbol{U}^\intercal\|_{F}^2/\|\mathcal{Y}\|_{F}^2$, where $\boldsymbol{U}$ is obtained from the SVD of basis evaluation matrices $\boldsymbol{\Phi}=\boldsymbol{U}\boldsymbol{D}\boldsymbol{V}^{\intercal}$ for a given $M_1,M_2$. This quantity can be considered as an estimate of the proportion of variance explained by $\text{span}(\boldsymbol{\phi}_{M}\otimes\boldsymbol{\phi}_{M})$.}
\par 
\noindent{{\bf \textit{Spherical Triangulation}}: The optimization problem \eqref{eqn:empirical_optm_reformulated_discretized_AO} is also dependent on the underlying spherical triangulation, denoted $\mathcal{T}_d$, used to define the spherical spline basis. We use a simple pruning heuristic to design $\mathcal{T}_d$, independently for each copy of $\mathbb{S}^2_d$, such that the local density of basis functions is aligned to the spatial distribution of $\boldsymbol{X}$. \markrevised{See the Supplemental Materials Section~\ref{ssec:triangulation_selection} for details.}

\section{Simulation Study}\label{sec:simulation}
 
\markrevised{In this section, we study several important aspects of our proposed procedure using simulated data. Specifically, the convergence of the proposed reduced-rank function space in $N$ and $K$ for both the in-sample error (residual) and the generalization error is considered in Section~\ref{ssec:recon_analysis}, Section~\ref{ssec:computational_perf} assesses the computational performance of Algorithm \ref{alg:rank_1_ssb}, and  Section~\ref{ssec:sim_inference} evaluates the local inference task presented in Section~\ref{ssec:inference} on a simulated two-group testing problem.}  All experiments were performed using MATLAB/2020 on a Linux OS with a 2.4 GHz Intel Xeon CPU E5-2695 and 32GB of RAM. 

\subsection{Reconstruction Error Analysis}\label{ssec:recon_analysis}
\markrevised{We simulate continuous connectivity according to:
$U(\omega_1, \omega_2) = \sum_{k=1}^{K_{true}} S_k\xi_k(\omega_1)\xi_k(\omega_2) $
where $\xi_k(\omega) = \boldsymbol{c}_{k}^\intercal \boldsymbol{\phi}_{M}(\omega)$, for $\boldsymbol{\phi}_{M}(\omega) = (\boldsymbol{\phi}_{M_{1}}^\intercal(\omega), \boldsymbol{\phi}_{M_{2}}^\intercal(\omega))^\intercal$ and $\boldsymbol{\phi}_{M_{d}}$ is the linear spherical spline basis formed by the Delaunay  triangulation of $M_1=M_2=410$ vertices. To generate the $\xi_k$, we independently sampled $\boldsymbol{c}_{k}\sim\mathcal{N}(\boldsymbol{0},0.2^2\boldsymbol{I})$. 
The random coefficients $S_{k}$ were drawn independently from $\mathcal{N}(0, k^{-1})$. The true rank was fixed to $K_{true}=20$. The $U$ were evaluated at all pairs of $n_1 + n_2 = 4,121$ points in the grid $\boldsymbol{X}$ over $\Omega$ discussed in Section~\ref{sec:data}.}
\par 
We study the connectivity reconstruction error as a function of $K$ for training sample sizes $N\in\{10,50,100\}$. For each sample size, in addition to the training data, we independently sampled  $100$ $U_i$'s to serve as a test set for estimating the generalization error of $P_{\mathcal{V}_{K}}$, for $\mathcal{V}_{K}$ constructed from the training set. For each setting, 100 replicated experiments were performed and the mean integrated square error (MSE) of the reconstructed connectivity was estimated for both training and testing data.  We used the marginal spline basis system for approximation and fixed a small roughness penalty parameter. The sparsity constraint was not active.
\par 
Figure~\ref{fig:simulation_analysis}a shows the MSE as a function of $K$ for different choices of $N$. As $N$ increases, we observe the in-sample (training) and out-sample (testing) error curves converge to one another. The black curve is $\propto(K+1)^{-1}$, in accord with the asymptotic error bound for the global optimum derived in Theorem~\ref{thrm:sample_convergence_rate}, with constant multiple defined by averaging the errors computed for the rank 1 case. We see that for all $K$ and $N$, on average, the errors obey the theory. Recall that the convergence of the proposed AO procedure is to a stationary point, which may not be the global solution. These simulation results indicate that the proposed AO procedure can construct solutions that exhibit the established convergence properties.
\begin{figure}[t]
    \centering
    \includegraphics[width=\textwidth]{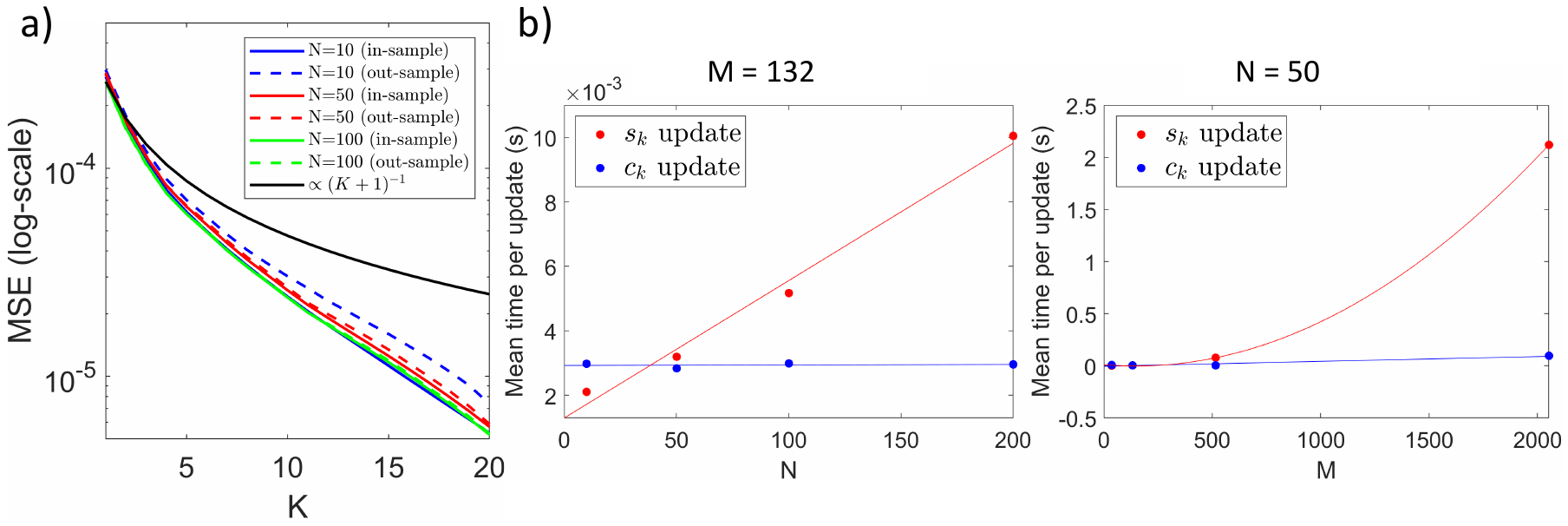}
    \caption{(a) Average sample (solid-lines) and generalization (dashed-lines) error as a function of model rank $K$ for several $N$. (b) Average per-iteration computational time for the updates \eqref{eqn:ctilde_update} and \eqref{eqn:s_update} as a function of sample size $N$ and marginal rank $M$.}
    \label{fig:simulation_analysis}
\end{figure} 

\subsection{Computational Performance}\label{ssec:computational_perf}
We now investigate the computational performance of the rank-1 updates in Algorithm~\ref{alg:rank_1_ssb}, which require iteratively solving \eqref{eqn:s_update} and \eqref{eqn:ctilde_update}. \markrevised{We use the same generative model for the connectivity as Section 6.1 and} consider all combinations of sample sizes $N\in\{10,50,100,200\}$ and marginal ranks $M\in\{36, 132, 516, 2052\}$. For each configuration, Algorithm~\ref{alg:rank_1_ssb} was run 10 times, and the mean per-update computational time was recorded along with the number of iterations required for convergence. Convergence was assessed by computing the relative change in the objective function between iterations of the inner loop in Algorithm~\ref{alg:rank_1_ssb}, terminating when a sufficiently small value ($10^{-6}$) was reached. The median number of iterations to convergence was $4$, with the minimum and maximum being $3$ and $14$, respectively, indicating that our algorithm converges quickly under a variety of settings.
\par 
Figure~\ref{fig:simulation_analysis}b displays the average computational time per update of both  $\boldsymbol{c}_k$ and $\boldsymbol{s}_k$. For updating $\boldsymbol{s}_k$, the computational time has an approximately linear relationship with the sample size $N$ (left panel) and an approximately quadratic relationship with spline rank $M$ (right panel). This is predicted by the theoretical complexity analysis: update  \eqref{eqn:s_update} can be computed at a computational cost of $O\left(N\left(M + M^2\right) + M^2\right)$. Updating  $\boldsymbol{c}_k$ is quite fast, on the order of $0.1$ seconds or less, for all the settings considered. 
\subsection{Two Group Continuous Sub-network Detection}\label{ssec:sim_inference}

\markrevised{In this section, we explore our method in the context of detecting local differences between two groups. Define two populations of functional data using the generative model:
$$
U_l(\omega_1, \omega_2) = \mu_{l}(\omega_1,\omega_2) + U(\omega_1,\omega_2),\qquad l \in \{1,2\}
$$
where $\mu_{l}$ is a population specific effect for the group $l$ and $U$ is a common random field. Specifically,
$$
\mu_{l}(\omega_1,\omega_2) := S_{0l}H_{h}((\omega_1,\omega_2)|(\omega_1^{*},\omega_2^{*}))\mathbb{I}\{d_{\mathbb{S}^2}(\omega_d,\omega_d^{*})<0.1,\text{ } d\in\{1,2\}\}
$$
with $H_{h}((\omega_1,\omega_2)|(\omega_1^{*},\omega_2^{*}))$ the symmetrized heat-kernel discussed in Section~\ref{ssec:subj_level_estim} with bandwidth $h=0.4$ centered at $(\omega_1^{*},\omega_2^{*})\in\Omega\times \Omega$, where $\omega_1^{*},\omega_2^{*}$ are sampled uniformly on $\Omega$. The random effect sizes are simulated according to $S_{0l}\sim\mathcal{N}(v_l, \eta^2)$, where 
$$
v_l = -\frac{v_0}{2}\mathbb{I}\{l=1\} + \frac{v_0}{2}\mathbb{I}\{l=2\},
$$
$\eta=0.1$ and $v_0$ is varied to simulate different signal sizes. The field $U$ is defined using the random sum of symmetric separable functions as outlined in Section~\ref{ssec:recon_analysis}, with the one change being the basis functions are determined according to: 
$\xi_k(\omega) = [\boldsymbol{a}_k\odot\boldsymbol{c}_{k}]^\intercal \boldsymbol{\phi}_{M}(\omega)$, for $\boldsymbol{a}_k\overset{iid}{\sim}\text{Bernoulli}(0.1)$ and $\odot$ is the Hadamard product, which is independently sampled to introduce sparsity. This set-up simulates the situation in which the difference between groups is confined within a small subset of connections. Our task is to be able to reliably identify the differential connections within a highly localized region.} 
\par 
\markrevised{We study the statistical power of our method under a variety of effect (signal) and sample sizes, $v_0\in\{0.005, 0.0075, 0.01, 0.0125, 0.025\}$ and $N=\{30, 50, 100\}$, respecitvely. Algorithm~\ref{alg:rank_1_ssb} is used to estimate $\mathcal{V}_{25}$ with locally supported elements $\xi_k$, where the sparsity level is automatically selected using the clustering approach discussed in Section~\ref{ssec:hyper_param_select}. Local inference is performed using the testing procedure outlined in Section~\ref{ssec:inference}, controlling for FCP at the $0.05$ level. For each simulation setting, we consider 100 Monte-Carlo (MC) replications.} 
\par
\markrevised{To quantify the detection performance, we computed the MC-average of the point-wise empirical coverage proportion (CP) and the empirical false coverage proportion (FCP), which can be defined here as
$$
    \text{CP} := \frac{1}{N_{mc}}\sum_{n=1}^{N_{mc}}\mathbb{I}\{(\omega_1^{*},\omega_2^{*})\in\mathcal{C}_n\}, \quad   \text{FCP}:= \frac{1}{N_{mc}}\sum_{n=1}^{N_{mc}}\mathbb{I}\{(\omega_1^{*},\omega_2^{*})\notin\mathcal{C}_n ,\mathcal{C}_n\neq\emptyset\},
$$
respectively, where $\mathcal{C}_{n}$ denotes continuous subnetwork cover formed using the $n$'th Monte-Carlo dataset, and $N_{mc} = 100$. 
To assess the specificity, we compute the empirical proportion of domain coverage (DC): 
$$
\text{DC}:=\sum_{n=1}^{N_{mc}}\left(\int_{\mathcal{C}_{n}}\mathbb{I}\{\mathcal{C}_n\neq \emptyset\}d\omega_1d\omega_2/(4(4\pi)^2)\right)\Big/\sum_{n=1}^{N_{mc}}\mathbb{I}\{\mathcal{C}_n\neq \emptyset\},
$$
which measures the average size of the $\mathcal{C}_{n}$'s relative to the total area of $\Omega\times \Omega$ ($4(4\pi)^2$). For a baseline comparison, we use permutations to test edgewise mean differences along with \cite{benjamini1995} correction to control FDR $\le 0.05$. We also compare our method to the popular Network-based statistic (NBS) \citep{zalesky2010_nbs} approach to identify significant sub-networks. We use the python implementation of NBS from the \textbf{brainconn} package.}
\par 
\markrevised{The top left panel of Figure~\ref{fig:subnetwork_detection_analysis} plots the estimated coverage proportion of our procedure as a function of $N$ for each $v_0$ considered. The statistical power, as quantified by the coverage proportion (CP), increases with $N$ and decreases for smaller effect size $v_0$, as expected. The results displayed in the middle plot of Figure~\ref{fig:subnetwork_detection_analysis} show that the testing procedure controls the FCP at the 0.05 level for most scenarios studied, except for the low signal, low sample ($v_0=0.005/0.0075, N=30$) cases, where the FPCs are only slightly above at 0.06. The FCP decreases with increasing sample size for all signal sizes considered. The right plot of Figure~\ref{fig:subnetwork_detection_analysis} shows the domain coverage (DC) as a function of $N$. The subnetwork cover tends to shrink (becoming more precise) with increasing $N$. Coupling this result with the positive relationship between coverage proportion and sample size indicates that our inference procedure displays desirable large sample behavior, generating increasingly tight subnetwork covers that cover the true simulated significant edge $(\omega_{1}^{*}, \omega_{2}^{*})$ with high probability.}
\par 
\markrevised{Edgewise testing with FDR correction resulted in no significant discoveries for all cases considered. This is not unexpected, as the number of tests is $\propto (n_1+n_2)^2$, rendering independent edgewise testing infeasible. For NBS, we could not run the entire MC study due to exceedingly long computational time (a single run with $v_0=0.1$ and $N=100$ took $\approx 6.5$ days). Our partial results show that even in this high-signal, large sample size setting, NBS could not identify significant connections. The observed computational bottleneck is not unexpected, as inference using the NBS algorithm comes at a computational cost of $O\left(P\left(n_1+n_2\right)\left(1+n_1+n_2\right)\right)$ ($P$ is the number of permutations), and thus scales poorly with the grid size. In comparison, the cost of local inference using our method is driven by the iterative updates \eqref{eqn:ctilde_update} and \eqref{eqn:s_update}, which are crucially independent of $n_1+n_2$, with only trivial additional computational expenditure for permutations on the coefficients of the $\xi_k$'s.}

\begin{figure}
    \centering
    \includegraphics[width=\textwidth]{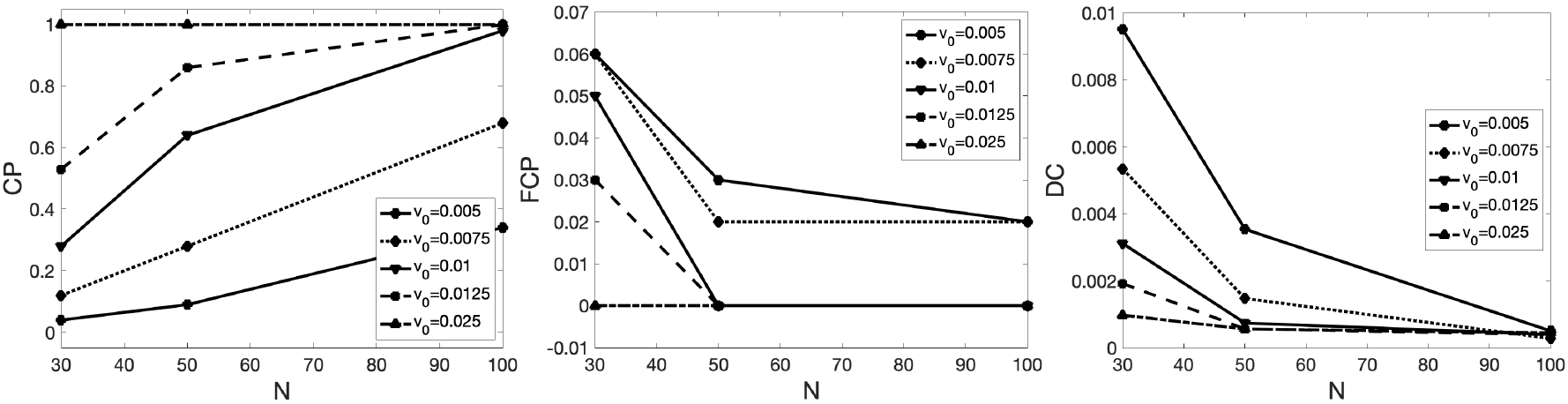}
    \caption{\markrevised{From left to right: Empirical coverage proportion (CP), false coverage proportion (FCP) and domain coverage (DC), estimated from the Monte-Carlo simulation results.}}
\label{fig:subnetwork_detection_analysis}
\end{figure}

\section{Real Data Applications}\label{sec:real_data_experiments}
\markrevised{In this section, we showcase our reduced-rank continuous connectivity representation framework—referred to here as CC—by applying the three main statistical procedures outlined in Section~\ref{ssec:inference} to the HCP structural connectivity data discussed in Section~\ref{sec:data}.}
We used the computation grid $\boldsymbol{X}$ on $\Omega$ with $n_1+n_2 = 4,121$, as outlined in Section~\ref{sec:data}. To form $\widehat{U}_i$ from the streamline endpoints of subject $i$ at all points on $\boldsymbol{X}$, the KDE bandwidth was set to be $h=0.005$ in Equation~\eqref{eqn:KDE_cc}, in accordance with the experiments in \cite{cole2021}. For all models considered in this section, we used a spherical spline basis with $M_1=M_2=410$, selected using an $85\%$ threshold on the criteria described in Section~\ref{ssec:hyper_param_select}. 
The roughness penalty parameter was selected to be $\alpha_1=10^{-8}$. The sparsity constraint was not active in the analysis in Sections \ref{ssec:reproducibility_main} and \ref{ssec:compare_network_embed},  while in the analysis in Section \ref{ssec:network_discovery} we selected \markrevised{$n_{\alpha_{2}}$ using the automated clustering approach discussed in Section~\ref{ssec:hyper_param_select}}. 

\subsection{CC Reproducibility Analysis}\label{ssec:reproducibility_main}
\markrevised{Given the numerous uncertainties in the brain imaging pipeline, it is crucial to evaluate the reproducibility of any neuroimaging analysis method. Details regarding the evaluations of our method's reproducibility are available in Section~\ref{ssec:reproducibility} of the Supplemental Material. Overall, our method exhibits excellent reproducibility across both scans (i.e., same subjects scanned in two separate sessions) and mesh sizes (i.e., the same subjects connectivity rendered at both 80k and 5k grid points).}  

\subsection{Relating Brain Networks to Traits}\label{ssec:compare_network_embed}

We compare the embeddings produced by our continuous approach to those constructed from atlas-based networks using several state-of-the-art joint network embedding methods, namely, Tensor-Network PCA (TN-PCA) \citep{zhang2019}, Multiple Random Dot-Product Graph model (MRDPG) \citep{nielsen2018} and Multiple Adjacency Spectral Embedding method (MASE) \citep{arroyo2021b}, on both hypothesis testing and prediction tasks. \markrevised{As a baseline, we also consider the approach of using the off-diagonal elements of the connectome matrices directly (OffDiag).} For evaluation, we used the sample of 437 female HCP subjects and their corresponding traits. 
\par 
Streamline count-based connectome matrices were formed from each subject's tractography result using both the Desikan (with 68 cortical parcels) and Destriex (with 148 cortical parcels) atlases. MRDPG and MASE require a binary adjacency matrix representation of the connectivity, which was obtained by thresholding the count-based connectome matrices, while TN-PCA was applied directly to the count matrices. To facilitate fair comparison, $K=100$ embedding dimensions were used for all methods, except MRDPG applied to the Desikan atlas, since this method requires $K\le V$, and hence we let $K=68$. The K-dimensional embeddings  were used as representations of the SC in subsequent analysis tasks.  

\subsubsection{Hypothesis Testing}\label{sssec:hypothesis_testing}
We first compare the power of the embeddings produced by different methods for identifying group differences. The groups are defined as follows: for each of the 80 measured traits in the categories \textit{cognition}, \textit{emotion}, and \textit{sensory}, two groups were created by selecting the top 100 and bottom 100 of the 437 HCP females, in terms of their measured trait score. For each trait, we use the MMD test \citep{gretton2012kernel} to test the null hypothesis that the groups of embedding vectors were drawn from the same distribution. The corresponding p-values were computed using 10,000 Monte-Carlo resampling iterations and corrected for false discovery rate (FDR) control using \cite{benjamini1995}.
\par 
\markrevised{The five panels in Figure~\ref{fig:global_inference}a show the p-value results for different embedding methods. The discrete network-based analysis was performed using the Destriex atlas. The y-axis gives the negative log transformed (raw) p-values and the colors indicate significant discoveries under a couple FDR control levels. With a threshold of $\text{FDR}\le 0.05$, the embeddings produced by our method are able to identify 22 significant discoveries, compared to 7 or less for the competitors. In Figure~\ref{fig:global_inference}b, we see that the empirical CDF of the p-values for our method has the largest departure from uniformity, the distribution expected under the null hypothesis of no difference.}
\par 
\markrevised{Figure~\ref{fig:global_inference_Desikan} in Supplemental Section~\ref{ssec:brain_network_traits} shows the same analysis for the network-based approaches using the Desikan atlas. The results show both fewer, and perhaps more saliently, different discoveries than using the Destriex atlas, illustrating the sensitivity of the discrete approaches to the choice of atlas. Not only does our continuous framework produce uniformly more powerful embeddings by modeling the SC data at much higher resolutions, but also the atlas-independence reduces the contingency of the reported results.} 

\begin{figure}[t]
    \centering
    \includegraphics[scale=0.58]{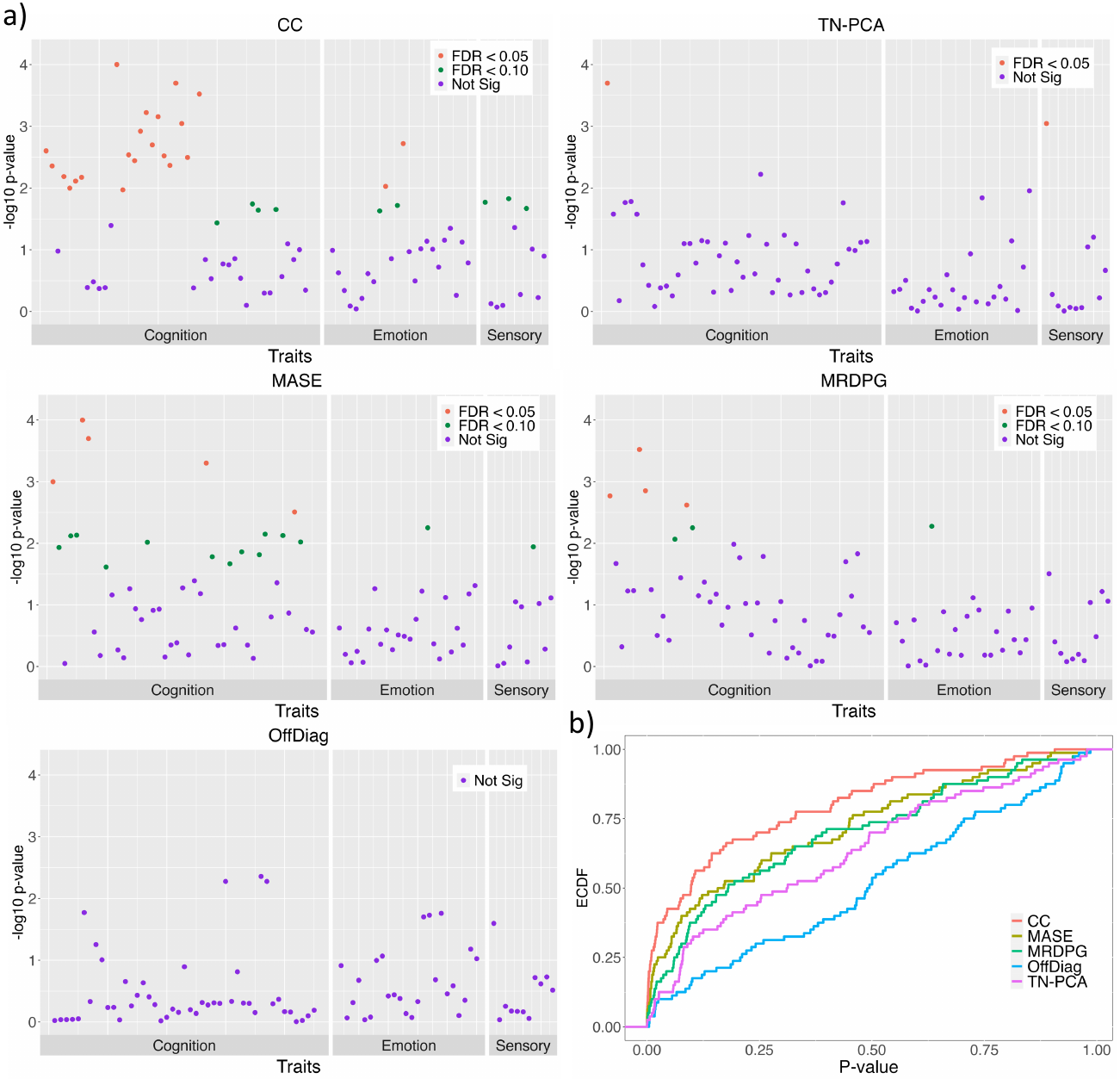}
    \caption{\markrevised{Results from the global hypothesis testing task using the Destriex atlas. a) $-\text{log}_{10}$-transformed uncorrected p-values from the MMD test for all 80 traits. b) Empirical cumulative distribution functions (ECDFs) of the p-values for each analysis method.}}
    \label{fig:global_inference}
\end{figure}

\subsubsection{Trait Prediction}\label{sssec:trait_prediction}
We now compare the performance of the different methods for the task of predicting various trait measurements from the SC. Many of the trait measurements in the HCP are under the same category (e.g., fluid intelligence, executive function and emotion recognition) and can be highly correlated with one another. Therefore, we constructed a set of composite measures using principal component analysis (PCA). Specifically, we first grouped these measurements, discarding any that were binary or categorical. The categories considered were: fluid intelligence, processing speed, self-regulation and impulse control, sustained attention, executive function, psychiatric, taste, emotional recognition, anger and hostility and finally BMI and weight. A PCA was performed for each category and the first $n$ PCs that explain $\ge 90\%$ of the variance were selected. Each subject's measurements were then projected onto each of the principal directions to get PC scores, which were used as outcomes for the trait prediction task. In all, there were 31 of these.
\par 
The embeddings were used as features in a LASSO regression model for the various outcomes of interest. For each outcome of interest, we performed 100 random 80-20 train-test splits of the data. The LASSO regularization parameter was selected with cross-validation using the training data. The trained model was then applied to the test set and the Pearson correlation ($\rho$) between the predicted and observed values was recorded. Any outcome whose predictions were not deemed significantly associated (random train-test split average $\rho > 0.1$) with \textit{any} of the embedding frameworks was discarded from the analysis. In all, 15 out of the original 31 outcomes met this criterion. 
\par 
Table~\ref{tab:concon_vs_atlas} records the Pearson correlation averaged over the random train-test splits for the significant outcomes for each method. The outcomes are labeled with the convention \textit{assessment}-\textit{PC number}. The discrete networks used by the competitor methods were formed with the Destriex atlas.  Our method outperforms all other methods for 12 out of the 15 outcomes. In several cases, the improvement is dramatic, e.g. Anger and Hostility-2, Emotion Recognition-4,  Executive Function-2. This suggests potentially large gains in predictive capability when analyzing connectivity data at very high resolutions. Table~\ref{tab:desikan_prediction_results} in Supplemental Section~\ref{ssec:brain_network_traits} gives the corresponding results for the networks formed using the Desikan atlas. Echoing the results of Section~\ref{sssec:hypothesis_testing}, for many cases, we note the predictions using the discrete network embedding techniques are highly sensitive to the choice of atlas.
\begin{table}[t]
\centering
\tiny
\begin{tabular}{r|rrrrrr}
\toprule
 &  \multicolumn{4}{c}{\textbf{Method}} \\
  Outcome & CC & TN-PCA & MASE & MRDPG & OffDiag\\
\midrule
Fluid Intelligence-1 & \textbf{0.1571} (0.0076) & 0.1122 (0.0069) & 0.1321 (0.0077) & 0.1235 (0.0084) & 0.1146 (0.0086) \\
Self Regulation-1 & \textbf{0.2083} (0.0083) & -0.0460 (0.0090) & 0.1280 (0.0081) & 0.1410 (0.0086) & 0.0347 (0.0089) \\
Self Regulation-3 & 0.0376 (0.0092) & 0.0464 (0.0103) & -0.0117 (0.0081) & \textbf{0.1289} (0.0097) & 0.0264 (0.0086) \\
Sustained Attention-2 & \textbf{0.1251} (0.0096) & -0.0412 (0.0084) & -0.0343 (0.0083) & -0.0031 (0.0092) & -0.0071 (0.0062) \\
Executive Function-1 & \textbf{0.1464} (0.0100) & 0.1072 (0.0079) & 0.0415 (0.0079) & 0.0881 (0.0075) & 0.1025 (0.0082) \\
Executive Function-2 & \textbf{0.1504} (0.0097) & -0.0116 (0.0092) & -0.0631 (0.0086) & 0.0521 (0.0102) & 0.0929 (0.0110) \\
Psychiatric-3 & \textbf{0.1070} (0.0088) & -0.0470 (0.0080) & -0.0107 (0.0103) & -0.0274 (0.0079) & -0.0276 (0.0093) \\
Psychiatric-6 & -0.0441 (0.0087) & \textbf{0.1277} (0.0086) & 0.1001 (0.0085) & 0.1239 (0.0096) & 0.0171 (0.0102) \\
Taste-1 & \textbf{0.1331} (0.0087) & 0.0164 (0.0087) & 0.0466 (0.0080) & 0.0214 (0.0087) & 0.0459 (0.0105) \\
Emotion Recognition-1 & \textbf{0.1023} (0.0093) & 0.0895 (0.0091) & 0.0896 (0.0080) & 0.0160 (0.0083) & 0.0378 (0.0131) \\
Emotion Recognition-2 & 0.0113 (0.0094) & 0.0489 (0.0089) & \textbf{0.1145} (0.0083) & -0.0436 (0.0092) & -0.0391 (0.0083) \\
Emotion Recognition-4 & \textbf{0.1428} (0.0084) & -0.0329 (0.0084) & -0.0659 (0.0090) & -0.0349 (0.0102) & -0.0190 (0.0051) \\
Anger and Hostility-2 & \textbf{0.2111} (0.0076) & 0.0011 (0.0096) & -0.0369 (0.0095) & -0.0379 (0.0081) & 0.0908 (0.0102) \\
Anger and Hostility-3 & \textbf{0.2062} (0.0094) & -0.0078 (0.0092) & 0.1850 (0.0090) & 0.0576 (0.0082) & -0.0057 (0.0084) \\
BMI and Weight-1 & \textbf{0.3153} (0.0076) & 0.2365 (0.0090) & 0.0866 (0.0085) & 0.1519 (0.0092) & 0.2037 (0.0089) \\
\hline
\end{tabular}
\caption{\markrevised{Trait prediction results for our continuous embeddings, denoted CC, along with TN-PCA, MASE, MRDPG and OffDiag using the Destriex atlas. The Pearson correlation between the true and predicted outcomes was averaged over the 100 random train-test splits. Standard errors are shown in parentheses.}}
\label{tab:concon_vs_atlas}
\end{table}

\subsection{Continuous Subnetwork Discovery}\label{ssec:network_discovery}

In this section, we illustrate how to identify connections that are different between groups using our continuous framework.  In order to focus the analysis, we consider the cognitive trait impulsivity, as measured by \textit{delay discounting}. Delay discounting measures the tendency for people to prefer smaller, immediate rewards over larger, but delayed rewards. It is known that steep discounting is associated with a variety of psychiatric conditions, including antisocial personality disorders, drug abuse and pathological gambling \citep{odum2011}. The HCP collects several measures of delay discounting on each participant as a part of a self-regulation/impulsivity assessment. In this study we focus on one of the measures, namely, the subjective value of $\$200$ in 6 months. For this measure, a series of trials are performed in which the subjects are made to choose between two alternatives: $\$200$ in 6 months or a smaller amount today. For each trial, reward amounts are adjusted based on the subject's choices to iteratively determine an indifference amount that identifies the participants subjective value of $\$200$ in 6 months. For more information on how this measurement is collected, we refer the interested reader to \cite{estle2006}.
\par 
\begin{figure}[!ht]
    \centering
    \includegraphics[scale=0.75]{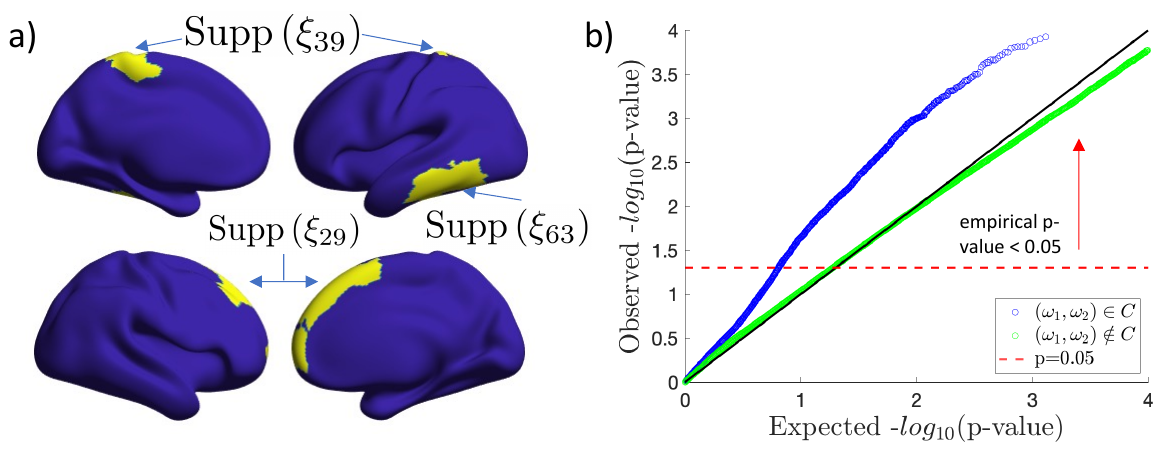}
    \caption{\markrevised{(a) The support sets (yellow) of the basis functions found to be significantly associated with connectivity differences in the high vs. low delay discounting groups. (b) QQ-plots for p-values from t-tests at all points $(\omega_1,\omega_2)\in \boldsymbol{X}\times\boldsymbol{X}$, colored by whether $(\omega_1,\omega_2)$ was contained within continuous subnetwork cover.}
    }
    \label{fig:subnet_select}
\end{figure}
From the sample of 437 female HCP subjects, we sub-selected and classified them to have high or low discounting, according to their subjective value of $\$200$ at 6 months. In particular, the subject was classified to the high discounting group if their subjective value $\le \$40$, and to the low discounting group if it was  $\ge\$160$. This resulted in a total sample size of 142: 64 in the high discounting group and 78 in the low discounting group. \markrevised{Locally supported basis functions $\mathcal{V}_{K}$ ($K=100$) were estimated using the full set of 437 HCP female subjects, with sparsity threshold selected using the automated clustering approach outlined in Section~\ref{ssec:hyper_param_select}. The local inference procedure formulated in Section~\ref{ssec:inference} was then applied to form $\mathcal{C}_{\alpha}$ with $\alpha=0.05$ using $10,000$ permutations.}
\par 
\markrevised{The coefficients associated with basis functions \markrevised{$\xi_{29},\xi_{39}$ and $\xi_{63}$} were identified to be significantly different between high and low discounting groups. Figure~\ref{fig:subnet_select}a shows the support sets of these selected basis functions (yellow regions) plotted on the cortical surface. To validate this finding, we conducted point-wise t-tests for all connections $\boldsymbol{X}\times\boldsymbol{X}$, and plotted the empirical p-values  vs.  the expected p-values (under $H_0$)  in Figure~\ref{fig:subnet_select}b, for both $(\omega_1,\omega_2)\in\mathcal{C}$ and $(\omega_1,\omega_2)\notin\mathcal{C}$. The skew of the -log$_{10}$-transformed p-values of the former indicates the presence of sub-regions of $\mathcal{C}$ where the continuous connectivity differs between groups.} 
\par 
\markrevised{The support sets show significant coverage of regions and connections in the prefrontal cortex. These results are strongly supported by the literature. Generally, areas in the prefrontal cortex are known to be critical for advanced decisions involving reward \citep{rogers1999}. Particular areas covered include the right superior frontal, middle frontal and orbitofrontal cortex. Previous studies have shown delay discounting to be associated with the grey matter volume in these areas \citep{wang2016_neuro,owens2017}. We also notice substantial coverage in the left paracentral region, which is thought to be primarily involved with motor and somatosensory processing, though task fMRI studies have identified this region to be active when choosing immediate rewards vs. control choices in delay discounting tasks \citep{stanger2013}. The final area of significant coverage is the rather large area on the left temporal lobe, a region which is known to be heavily involved with temporal information processing \citep{elias1999}. Additionally, \cite{olson2009} found strong evidence of significant alterations in white matter in this region between high and low discount groups. Recalling that the continuous subnetwork covers significant connections between any pair of points, all identified regions can be interpreted jointly in terms of possible subnetworks between them. The task fMRI literature has identified possible functional networks related to delay discounting which include areas in the prefrontal, parietal and temporal regions \citep{olson2009}. Such networks are consistent with the coverage implied by $\mathcal{C}$, suggesting a possible link between the significant functional and structural networks of importance.}

\section{Conclusion and Future Directions}\label{sec:conclusion}
This work introduces a novel modeling framework for the analysis of structural connectivity. We define the continuous structural connectivity as a latent random function on the symmetric product space of the cortical surface that governs the distribution of white matter fiber tract endpoints. This continuous representation allows us to bypass issues that plague the traditional discrete network based approaches but also poses new challenges for computation and modeling due to the super-high dimensionalities of the data. To facilitate tractable representation and analysis, we formulate a data-adaptive reduced-rank function space that avoids the curse of dimensionality. To construct a set of basis functions that span this space, we derive a penalized optimization problem and propose a novel computationally efficient algorithm for estimation. The proposed method was applied to several critical neuroscience applications, including hypothesis testing and trait prediction. Through comparison with state-of-the-art atlas-based network analysis methods, we demonstrated the superior performance of the proposed framework in real data analysis of HCP data.
\par 
We conclude by noting several possible extensions of the proposed framework. 
First, a unified method that is able to estimate $\mathcal{V}_K$ directly from the streamline endpoints without the need for the intermediate KDE smoothing step is desirable.
\markrevised{Second, forming the low rank model by decomposing the variability directly in the linear $L^2(\Omega\times\Omega)$ space may not be ideal, as this does not explicitly respect the constraints on $U$. There are several geometric frameworks for performing functional data modeling of random densities, e.g. \cite{srivastava2011,petersen2016}, but their extension to our case is non-trivial due to the complex multidimensional domain $\Omega\times\Omega$, and hence more investigation is required. 
Third, our method employs the Holm procedure to control the FWER when comparing random coefficients ($S_{lk}$) from two subject groups to ensure the control over the FCP of the random intensity functions. While the Holm procedure ensures tight control of the FWER regardless of the dependence structure of the test statistics, this versatility may lead to over-conservativeness in practice. Exploring alternative FWER control methods, such as those assuming mild conditions on the joint distribution of test statistics \citep{meinshausen2011}, could lead to a more powerful local inference procedure, and is therefore an important direction for future work.}

\par\bigskip
\textbf{\large Code}: Implementation of our method along with code for processing the SC data has been made publicly available: {\small \textbf{https://github.com/sbci-brain/SBCI\_Modeling\_FPCA}}
\bibliographystyle{chicago}
\bibliography{references}
\clearpage
\pagebreak

\begin{center}
  \textbf{\large Supplementary Material} 
\end{center}

\setcounter{equation}{0}
\setcounter{figure}{0}
\setcounter{table}{0}
\setcounter{section}{0}
\setcounter{page}{1}
\makeatletter
\renewcommand{\theequation}{S\arabic{equation}}
\renewcommand{\thefigure}{S\arabic{figure}}
\renewcommand{\thetable}{S\arabic{table}}
\renewcommand{\thesection}{S\arabic{section}}
\renewcommand{\bibnumfmt}[1]{[S#1]}
\renewcommand{\citenumfont}[1]{S#1}
\renewcommand{\theequation}{S.\arabic{equation}}
\renewcommand{\thesection}{S\arabic{section}}
\renewcommand{\thesubsection}{S\arabic{section}.\arabic{subsection}}
\renewcommand{\thetable}{S\arabic{table}}
\renewcommand{\thefigure}{S\arabic{figure}}
\renewcommand{\thetheorem}{S\arabic{theorem}}
\renewcommand{\theproposition}{S\arabic{proposition}}
\renewcommand{\thelemma}{S\arabic{lemma}}
\renewcommand{\theassumption}{S\arabic{assumption}}

\section{Theory and Proofs}\label{sec:theory_proofs}

\subsection{Proof of Theorem~\ref{thrm:sample_convergence_rate}}
In order to proceed, we make the following assumptions related to the smoothness of $U$.
\begin{definition}\label{defn:smoothness_class}
Let $K\in\mathbb{N}_{\infty}$ and $0 < B < \infty$. Define $L_{1,B}$ as the class of functions
$$
L_{1,B}(\mathcal{V}_K) := \{u \in \mathcal{H}: u = \sum_{k=1}^K s_k\xi_k\otimes\xi_k, \sum_{k=1}^K |s_k| < B\}.
$$ 
where $s_{k} := \langle u, \xi_k\otimes \xi_k\rangle_{L^2(\Omega\times\Omega)}$, for $\xi_k\otimes\xi_k \in \mathcal{V}_K$.
\end{definition}
\begin{assumption}[Eigenfunction Smoothness]\label{assm:smoothness_class}
There exists some $\mathcal{V}_K^{*}$ for $K\in\mathbb{N}_{\infty}$ and 
$0 < B_1 < \infty$, such that $\psi_k \in L_{1,B_{1}}(\mathcal{V}_K^{*})$ for all $k=1,2,...$.
\end{assumption}
\begin{assumption}[Moment Bound]\label{assm:moment_bound}
For some $0 < B_2 < \infty$, $\sum_{k=1}^\infty |Z_k| < B_2$, holds with probability 1. 
\end{assumption}
\begin{assumption}[Tail Decay]\label{assm:decay_rate}
$\sum_{k=1}^\infty \sqrt{\rho_k} < \infty$
\end{assumption}
The function class in Definition~\ref{defn:smoothness_class} has been used in the literature for establishing error convergence rates for deterministic function approximation algorithms \citeSupp{Temlyakov2003NonlinearMO,barron2008}. This assumption is closely related to the smoothness of the $\psi_k$'s, see the Section~\ref{ssec:eigenfunction_smoothness} for more details. Assumptions~\ref{assm:moment_bound} and \ref{assm:decay_rate} control the tail behavior of $U$. Note that Assumption~\ref{assm:decay_rate} introduces a slightly stronger condition on the decay rate of the spectrum than the one guaranteed by the mean square integrability of $C$, i.e. $\sum_{k=1}^\infty \rho_k <\infty$, though it is satisfied for many commonly used covariance kernels. 

\begin{proposition}\label{prop:as_smoothness}
Under Assumptions~\ref{assm:smoothness_class} and \ref{assm:moment_bound}, $U\in L_{1,B_1B_2}(\mathcal{V}_K^{*})$ almost surely. 
\end{proposition}
\begin{proof}
Notice that 
$$
\begin{aligned}
     U &= \sum_{k=1}^\infty Z_k \psi_k = \sum_{k=1}^\infty Z_k\sum_{j=1}^\infty c_{kj}\xi_j\otimes\xi_j \\ 
     &= \sum_{j=1}^\infty \left(\sum_{k=1}^\infty Z_k c_{kj}\right)\xi_j\otimes\xi_j
    \end{aligned},
$$
where $c_{kj} = \langle \psi_k, \xi_j\otimes\xi_j\rangle_{L^{2}(\Omega\times\Omega)}$. Then
$$
    \sum_{j=1}^\infty |\sum_{k=1}^\infty Z_k c_{kj}| \le  \sum_{j=1}^\infty\sum_{k=1}^\infty|Z_k||c_{kj}| \le B_1 \sum_{k=1}^\infty|Z_k| \le B_1B_2,
$$
where the last inequality holds almost surely. 
\end{proof}
\begin{proposition}\label{prop:Sj_bound}
Under Assumptions~\ref{assm:smoothness_class}, \ref{assm:moment_bound} and \ref{assm:decay_rate}:
$$
\sum_{j=1}^\infty\mathbb{E}\left[S_{j}^2\right]^{1/2} \le  B_1\sum_{k=1}^\infty\sqrt{\rho_k},
$$
where $S_{j} = \langle U, \xi_{j}\otimes\xi_{j}\rangle_{L^{2}(\Omega\times\Omega)}$ for $\xi_j \in \mathcal{V}_K^{*}$.
\end{proposition}
\begin{proof}
Denote $c_{kj} = \langle \psi_k, \xi_j\otimes\xi_j\rangle_{L^{2}(\Omega\times\Omega)}$, we have 
\begin{equation}\label{eqn:bound_4_sj}
\begin{aligned}
     S_{j} &= \langle U, \xi_{j}\otimes\xi_{j}\rangle_{L^{2}(\Omega\times\Omega)} =  \langle \sum_{k=1}^\infty Z_k\psi_k, \xi_{j}\otimes\xi_{j}\rangle_{L^{2}(\Omega\times\Omega)} \\
     &= \sum_{k=1}^\infty Z_k\langle\psi_k, \xi_{j}\otimes\xi_{j}\rangle_{L^{2}(\Omega\times\Omega)} = \sum_{k=1}^\infty Z_k \langle\sum_{l=1}^\infty c_{lk}\xi_{l}\otimes\xi_{l}, \xi_{j}\otimes\xi_{j}\rangle_{L^{2}(\Omega\times\Omega)} \\
     & = \sum_{k=1}^\infty Z_k \sum_{l=1}^\infty c_{lk} \langle \xi_{l}\otimes\xi_{l}, \xi_{j}\otimes\xi_{j}\rangle_{L^{2}(\Omega\times\Omega)}  = \sum_{k=1}^\infty Z_k \sum_{l=1}^\infty c_{lk}\delta_{lj} \\
     & = \sum_{k=1}^\infty Z_kc_{kj} \\
\end{aligned}
\end{equation}
and 
$$
\begin{aligned}
     \mathbb{E}\left[ S_{j}^2\right] =  \mathbb{E}\left[\sum_{k=1}^\infty Z_kc_{kj}\sum_{l=1}^\infty Z_lc_{lj}\right] =  \sum_{k=1}^\infty\mathbb{E}\left[ Z_k^2\right]c_{kj}^2
\end{aligned}
$$
so 
$$
\begin{aligned}
     \sum_{j=1}^\infty\mathbb{E}\left[S_{j}^2\right]^{1/2} &= \sum_{j=1}^\infty\left[\sum_{k=1}^\infty\mathbb{E}\left[ Z_k^2\right]c_{kj}^2\right]^{1/2} \\
     &\le \sum_{j=1}^\infty\sum_{k=1}^\infty\left[\mathbb{E}\left[ Z_k^2\right]c_{kj}^2\right]^{1/2} = \sum_{j=1}^\infty\sum_{k=1}^\infty\sqrt{\rho_k}|c_{kj}| \\
     &= \sum_{k=1}^\infty\sqrt{\rho_k}\sum_{j=1}^\infty |c_{kj}| \le \sum_{k=1}^\infty\sqrt{\rho_k} B_1
\end{aligned}
$$
\end{proof}

\noindent{\textbf{Proof of Theorem~\ref{thrm:sample_convergence_rate}}}
\begin{proof} 
Notice that we can write
\begin{equation}\label{eqn:reccurence}
\begin{aligned}
    \left\|R_{k,i}\right\|_{L^{2}}^2 &= \left\| U_i - P_{k}(U_i)\right\|_{L^{2}}^2 \\
    &= \left\| U_i - P_{k-1}(U_i) - \langle U_i - P_{k-1}(U_i), \xi_k\otimes\xi_k\rangle \xi_k\otimes\xi_k\right\|_{L^{2}}^2  \\
    &= \left\| U_i - P_{k-1}(U_i)\right\|_{L^{2}}^2 - |\left\langle U_i - P_{k-1}(U_i), \xi_k\otimes\xi_k\right\rangle_{L^{2}}|^2 \\
    & \equiv \left\| R_{k-1,i}\right\|_{L^{2}}^2 - |\left\langle R_{k-1,i}, \xi_k\otimes\xi_k\right\rangle_{L^{2}}|^2 
\end{aligned}
\end{equation}
For clarity, denote $S_{i,j} = \langle U_i, \xi_j\otimes\xi_j\rangle_{L^{2}}$ with $\xi_j \in \mathcal{V}_K^{*}$. We have, 
\begin{equation}\label{eqn:bound_prev_reccurence}
\begin{aligned}
    \left\| R_{k-1,i}\right\|_{L^{2}}^2 &=  \left\langle R_{k-1,i}, R_{k-1,i}\right\rangle_{L^{2}} \\ 
    &= \left\langle R_{k-1,i}, U_i\right\rangle_{L^{2}} \\
    &= \left\langle R_{k-1,i}, \sum_{j = 1}^\infty S_{i,j}\xi_{j}\otimes\xi_{j}\right\rangle_{L^{2}} \\
    &= \sum_{j=1}^\infty S_{i,j}\left\langle R_{k-1,i},  \xi_{j}\otimes\xi_{j}\right\rangle_{L^{2}}\\ 
    &\le \sum_{j=1}^\infty|S_{i,j}||\left\langle R_{k-1,i},  \xi_{j}\otimes\xi_{j}\right\rangle_{L^{2}}|.\\
\end{aligned}
\end{equation}
We have that 
\begin{equation}\label{eqn:cauchy_schwarz_bound}
\begin{aligned}
&N^{-1}\sum_{i=1}^N\sum_{j=1}^\infty|S_{i,j}||\left\langle R_{k-1,i},  \xi_j\otimes\xi_j\right\rangle_{L^{2}}| = \sum_{j=1}^\infty N^{-1}\sum_{i=1}^N|S_{i,j}||\left\langle R_{k-1,i},  \xi_j\otimes\xi_j\right\rangle_{L^{2}}| \\
&\le \sum_{j=1}^\infty \left[N^{-1}\sum_{i=1}^N|S_{i,j}|^2\right]^{1/2}\left[N^{-1}\sum_{i=1}^N|\left\langle R_{k-1,i},  \xi_j\otimes\xi_j\right\rangle_{L^{2}}|^2\right]^{1/2} \\
&\le \sum_{j=1}^\infty\left[N^{-1}\sum_{i=1}^N|S_{i,j}|^2\right]^{1/2}\left[\sup_{\xi\in\mathbb{S}^{\infty}(\Omega)}N^{-1}\sum_{i=1}^N|\left\langle R_{k-1,i},  \xi\otimes\xi\right\rangle_{L^{2}}|^2\right]^{1/2}\\
\end{aligned}
\end{equation}
Combining \eqref{eqn:cauchy_schwarz_bound} and \eqref{eqn:bound_prev_reccurence} gives the bound 
$$
 \left[N^{-1}\sum_{i=1}^N \left\| R_{k-1,i}\right\|_{L^{2}}^2\right]^2 \le \left[\sum_{j=1}^\infty\left[N^{-1}\sum_{i=1}^N|S_{i,j}|^2\right]^{1/2}\right]^2\sup_{\xi\in\mathbb{S}^{\infty}(\Omega)}N^{-1}\sum_{i=1}^N|\left\langle R_{k-1,i},  \xi\otimes\xi\right\rangle_{L^{2}}|^2.
$$
By the definition of our greedy selection algorithm, we have that
$$
\sup_{\xi\in\mathbb{S}^{\infty}(\Omega)}N^{-1}\sum_{i=1}^N|\left\langle R_{k-1,i},  \xi\otimes\xi\right\rangle_{L^{2}}|^2 = N^{-1}\sum_{i=1}^N|\left\langle R_{k-1,i},  \xi_k\otimes\xi_k\right\rangle_{L^{2}}|^2.
$$
Plugging these results into the recurrence relation \eqref{eqn:reccurence}, we obtain
\begin{equation}\label{eqn:recursive_bound}
\begin{aligned}
N^{-1}\sum_{i=1}^N \left\|R_{k,i}\right\|_{L^{2}}^2 &= N^{-1}\sum_{i=1}^N \left\| R_{k-1,i}\right\|_{L^{2}}^2 - N^{-1}\sum_{i=1}^N |\left\langle R_{k-1,i}, \xi_k\otimes\xi_k\right\rangle_{L^{2}}|^2 \\
&\le N^{-1}\sum_{i=1}^N\left\| R_{k-1,i}\right\|_{L^{2}}^2\left(1 - \frac{N^{-1}\sum_{i=1}^N \left\| R_{k-1,i}\right\|_{L^{2}}^2}{\left[\sum_{j=1}^\infty\left[N^{-1}\sum_{i=1}^N|S_{i,j}|^2\right]^{1/2}\right]^2}\right). \\
\end{aligned}
\end{equation}
From the strong law of large numbers, $N^{-1}\sum_{i=1}^N|S_{i,j}|^2\overset{a.s.}{\rightarrow}\mathbb{E}\left[S_{j}^2\right]$, the bound in proposition~\ref{prop:Sj_bound}, and the continuous mapping theorem, it follows that taking the large sample limit of \eqref{eqn:recursive_bound} gives 
\begin{equation}\label{eqn:empirical_recurrence_prop_bound}
\lim_{N\rightarrow\infty}N^{-1}\sum_{i=1}^N \left\|R_{k,i}\right\|_{L^{2}}^2\le \lim_{N\rightarrow\infty}N^{-1}\sum_{i=1}^N \left\| R_{k-1,i}\right\|_{L^{2}}^2\left(1 - \frac{\lim_{N\rightarrow\infty}N^{-1}\sum_{i=1}^N\left\| R_{k-1,i}\right\|_{L^{2}}^2}{ B_1^2\left(\sum_{k=1}^\infty\sqrt{\rho_k}\right)^2}\right)
\end{equation}
almost surely. Notice that
$$
1=\|\psi_k\|_{L^{2}} = \left(\sum_{j=1}^\infty|\langle \psi_k,\xi_j\otimes\xi_j\rangle_{L^{2}}|^2\right)^{1/2} 
\le \sum_{j=1}^\infty|\langle \psi_k,\xi_j\otimes\xi_j\rangle_{L^{2}}|\le B_{1},
$$
where the first inequality comes from the fact that the $\|\cdot\|_1$ norm dominates $\|\cdot\|_2$ in $l^{p}$ sequence spaces and the second from Assumption~\ref{assm:smoothness_class}. Hence, we have 
\begin{equation}\label{eqn:R0_bound}
\begin{aligned}
     \lim_{N\rightarrow\infty}N^{-1}\sum_{i=1}^N\left\|R_{0,i}\right\|_{L^{2}}^2 = \lim_{N\rightarrow\infty}N^{-1}\sum_{i=1}^N\left\|U_{i}\right\|_{L^{2}}^2 \overset{a.s.}{\rightarrow} \sum_{k=1}^\infty\rho_k \le B_1^2 \left( \sum_{k=1}^\infty\sqrt{\rho_k}\right)^2,
\end{aligned}
\end{equation}
almost surely. Hence the sequence $\{\lim_{N\rightarrow\infty}N^{-1}\sum_{i=1}^N\left\|R_{k,i}\right\|_{L^{2}}^2\}_{k}$ is decreasing (in $k$).
\par 
We now derive  the final result by induction. For clarity, denote 
$r_k := \lim_{N\rightarrow\infty}N^{-1}\sum_{i=1}^N\left\|R_{k,i}\right\|_{L^{2}}^2.$
For $k=0$, it follows from Equation~\ref{eqn:R0_bound} that $r_0\le B_1^2\left(\sum_{j=1}^\infty\sqrt{\rho_j}\right)^2$. 
Assume that $r_{k}\le \frac{ B_1^2\left(\sum_{j=1}^\infty\sqrt{\rho_j}\right)^2}{k+1}$ for any $k\ge 1$. Then if $r_{k} \le \frac{B_1^2\left(\sum_{j=1}^\infty\sqrt{\rho_j}\right)^2}{k+2}$, clearly $r_{k+1} \le \frac{B_1^2\left(\sum_{j=1}^\infty\sqrt{\rho_j}\right)^2}{k+2}$ since it is a decreasing sequence. If $r_{k} \ge \frac{B_1^2\left(\sum_{j=1}^\infty\sqrt{\rho_j}\right)^2}{k+2}$, then using this fact along with the recurrence \eqref{eqn:empirical_recurrence_prop_bound} and induction hypothesis, we have
$$
r_{k+1} \le \frac{B_1^2\left(\sum_{j=1}^\infty\sqrt{\rho_j}\right)^2}{k+1}\left(1 - \frac{1}{B_1^2\left(\sum_{j=1}^\infty\sqrt{\rho_j}\right)^2}\frac{B_1^2\left(\sum_{j=1}^\infty\sqrt{\rho_j}\right)^2}{k+2}\right)= \frac{B_1^2\left(\sum_{j=1}^\infty\sqrt{\rho_j}\right)^2}{k+2},
$$ 
establishing the desired result.
\end{proof}

\subsection{Proof of Theorem ~\ref{thrm:continuous_subnetwork_discovery}}

\begin{proof}
Assume that $I\neq\emptyset$ and take $k\in I$. Denote the marginal random function $U_{j,\omega_{1}}(\cdot) := U_j(\omega_{1}, \cdot)$. Assume that $U_{1,\omega_{1}}\overset{\text{dist}}{=}U_{2,\omega_{1}}$ for any $\omega_1\in\text{Supp}(\xi_k)$. By the continuous mapping theorem, this implies $\int_{\Omega} U_{1,\omega_{1}}(\omega)\xi_k(\omega)d\omega\overset{\text{dist}}{=}\int_{\Omega}U_{2,\omega_{1}}(\omega)\xi_k(\omega)d\omega$, which in turn implies $S_{1k} \overset{\text{dist}}{=} S_{2k}$, which is a contradiction under $k\in I$. Hence, it must be the case that $U_{1,\omega_{1}}\overset{\text{dist}}{\neq}U_{2,\omega_{1}}$ for some $\omega_1\in\text{Supp}(\xi_k)$, which implies $\exists \omega_2\in\Omega$ such that $U_1(\omega_1,\omega_2) \overset{\text{dist}}{\neq} U_2(\omega_1,\omega_2)$.
\par 
We now show that $\omega_2$ must be in $\bigcup_{j\in I}\text{Supp}(\xi_j)$. We proceed by contradiction. Assume $\omega_2 \in \Omega \setminus \bigcup_{j\in I}\text{Supp}(\xi_j)$. Let $I^c = \{1,...,K\}\setminus I$, then $U_1(\omega_1, \omega_2)\overset{\text{dist}}{\neq} U_2(\omega_1,\omega_2)$ implies  
$$
\begin{aligned}
     \underbrace{\sum_{k \in I} S_{1k}\xi_k(\omega_1)\xi_k(\omega_2)}_{\xi_k(\omega_2)=0} + \sum_{j \in I^{c}} S_{1j}\xi_j(\omega_1)\xi_j(\omega_2)&\overset{\text{dist}}{\neq} \underbrace{\sum_{k \in I} S_{2k}\xi_k(\omega_1)\xi_k(\omega_2)}_{\xi_k(\omega_2)=0}  + \sum_{j \in I^{c}} S_{2j}\xi_j(\omega_1)\xi_j(\omega_2) \\ 
     \Rightarrow\sum_{j \in I^{c}} S_{1j}\xi_j(\omega_1)\xi_j(\omega_2)&\overset{\text{dist}}{\neq}  \sum_{j \in I^{c}} S_{2j}\xi_j(\omega_1)\xi_j(\omega_2)
      \Rightarrow\sum_{j \in I^{c}} S_{1j} \overset{\text{dist}}{\neq}  \sum_{j \in I^{c}} S_{2j}\\
\end{aligned}
$$
which is a contradiction since $S_{1j} \overset{\text{dist}}{=} S_{2j}$ for all $j \in I^c$. Hence, since $(\omega_1,\omega_2)\in\mathcal{C}$, the desired result follows. 
\end{proof}
\subsection{Proof of Theorem~\ref{thm:FCP_control}}
\begin{proof}
We are interested in bounding the probability 
$$
\text{FCP}(\mathcal{C}_{\alpha}) := \mathbb{P}\left[\mathcal{C}_{\alpha}\neq \emptyset , \text{ and } U_1(\omega_1,\omega_2) \overset{\text{dist}}{=} U_2(\omega_1,\omega_2), \forall (\omega_1,\omega_2)\in \mathcal{C}_{\alpha}\right].
$$
We proceed by cases. First, assume that $\boldsymbol{S}_1\overset{\text{dist}}{=}\boldsymbol{S}_2$, with $\boldsymbol{S}_j = (S_{11},...,S_{1K})^{\intercal}$ for $j=1,2$. Under the rank-$K$ approximation  
$
U_j\approx P_{\mathcal{V}_{K}}(U_{j}) = \sum_{k=1}^K S_{jk}\xi_k\otimes\xi_k
$, this implies the point-wise condition 
$$
\sum_{k=1}^K S_{1k}\xi_k(\omega_1)\xi_k(\omega_2) \overset{\text{dist}}{=} \sum_{k=1}^K S_{2k}\xi_k(\omega_1)\xi_k(\omega_2), \forall (\omega_1,\omega_2)\in \Omega\times\Omega.
$$
Furthermore, $\boldsymbol{S}_1\overset{\text{dist}}{=}\boldsymbol{S}_2$ is equivalent to the condition $H_0^{k}$ is true $\forall k$. 
By definition of the FWER control, we have that our testing procedure satisfies:
$$
\alpha \ge \mathbb{P}\left[ \exists k \ni H^{k}_0 \text{ rejected}\Big|H^{k}_0 \text{ true }\forall k\right] \ge  \mathbb{P}\left[ \mathcal{C}_{\alpha}\neq \emptyset \Big|H^{k}_0 \text{ true }\forall k\right] = \mathbb{P}\left[ \mathcal{C}_{\alpha} \neq \emptyset \Big| \boldsymbol{S}_1\overset{\text{dist}}{=}\boldsymbol{S}_2\right].
$$
and hence, 
$$
\begin{aligned}
\text{FCP}(\mathcal{C}_{\alpha}) 
\le \mathbb{P}\left[ \mathcal{C}_{\alpha}\neq \emptyset \Big| \boldsymbol{S}_1\overset{\text{dist}}{=}\boldsymbol{S}_2\right] \le \alpha.
\end{aligned}
$$
Now consider the case $\boldsymbol{S}_1\overset{\text{dist}}{\neq}\boldsymbol{S}_2$.
Let $J = \{k: S_{1k}\overset{\text{dist}}{\neq}S_{2k}\}$ and define $I = \{k\in J: \text{supp}(\xi_k\otimes\xi_k)\bigcap\mathcal{C}_{\alpha}\}$. If $I= \emptyset$, then this implies that $\mathcal{C}_{\alpha}$ would be constructed by false rejections, and by the assumed strong control of the FWER correction procedure
\begin{equation}\label{eqn:strong_FWER_control}
\begin{aligned}
    \alpha  &\ge\mathbb{P}\left[ \exists k \ni H^{k}_0 \text{ rejected for }k\in \{1,...,K\} \backslash J \Big|H^{k}_0 \text{ true }\forall k\in \{1,...,K\} \backslash J \right] \\
    &\ge\mathbb{P}\left[ \mathcal{C}_{\alpha}\neq \emptyset \Big|H^{k}_0 \text{ true }\forall k\in \{1,...,K\} \backslash J \right]. 
\end{aligned}
\end{equation}
Furthermore, from the definition of $J$, we have 
$$
\sum_{k\in \{1,...,K\} \backslash J}
S_{1k}\xi_k(\omega_1)\xi_k(\omega_2) \overset{\text{dist}}{=} \sum_{k\in \{1,...,K\} \backslash J} S_{2k}\xi_k(\omega_1)\xi_k(\omega_2), \forall (\omega_1,\omega_2)\in \Omega\times\Omega,
$$
and from $I=\emptyset$, we have 
$$
\sum_{k\in J} S_{1k}\xi_k(\omega_1)\xi_k(\omega_2) = 0 = \sum_{k\in J} S_{2k}\xi_k(\omega_1)\xi_k(\omega_2), \forall (\omega_1,\omega_2)\in \mathcal{C}_{\alpha},
$$
so as a result
$$
\sum_{k=1}^K S_{1k}\xi_k(\omega_1)\xi_k(\omega_2) \overset{\text{dist}}{=} \sum_{k=1}^K S_{2k}\xi_k(\omega_1)\xi_k(\omega_2), \forall (\omega_1,\omega_2)\in \mathcal{C}_{\alpha}
$$
holds using the the local support property. Coupling this with Equation~\ref{eqn:strong_FWER_control}, it follows that $\text{FCP}(\mathcal{C}_{\alpha}) \le \alpha$ for this case as well. 
\par 
Now, the final case is $\boldsymbol{S}_1\overset{\text{dist}}{\neq}\boldsymbol{S}_2$ and $I\neq \emptyset$. Take any $k^{*}\in I$ and define the operator $\mathcal{T}_{k^{*}}(u) = \int_{\mathcal{C}_{\alpha}} u(\omega_1,\omega_2) \xi_{k^{*}}(\omega_1)\xi_{k^{*}}(\omega_2)d\omega_1d\omega_2$. Applying the continuous mapping theorem to the point-wise equality of distribution condition in \eqref{eqn:FCP}, we have that 
$$
\mathcal{T}_{k^{*}}\left(\sum_{k=1}^K S_{1k}\xi_k(\omega_1)\xi_k(\omega_2) \right)\overset{\text{dist}}{=} \mathcal{T}_{k^{*}}\left(\sum_{k=1}^K S_{2k}\xi_k(\omega_1)\xi_k(\omega_2)\right) \rightarrow S_{1k^{*}}\overset{\text{dist}}{=} S_{2k^{*}},
$$
which is a contradiction, hence $U_1(\omega_1,\omega_2) \overset{\text{dist}}{=} U_2(\omega_1,\omega_2), \forall (\omega_1,\omega_2)\in \mathcal{C}_{\alpha}$ cannot hold in this case and so $\text{FCP}(\mathcal{C}_{\alpha}) = 0\le \alpha$.
\end{proof}
\subsection{Eigenfunction Smoothness}\label{ssec:eigenfunction_smoothness}

In this section, we provide more detail about the implications of Assumption~\ref{assm:smoothness_class}. First, we note that this condition in trivially satisfied when the eigenfunctions are separable, as taking $\mathcal{V}^{*}_{\infty}$ to be the collection of eigenfunctions clearly ensures $\psi_k\in L_{1,1}(\mathcal{V}^{*}_{\infty})$ for all $k$. A sufficient, though not necessary, condition for eigenfunction separability is the covaraince function $C$ itself being separable. 
\par 
The following proposition ensures that each $\psi_k$ can be represented by a countable set of separable functions.
\begin{proposition}\label{prop:uap}
For any  $u\in\mathcal{H}$, there exists at least one countable set $\mathcal{V}_K$, possibly with $K=\infty$, such that  $u\in\text{span}\left(\mathcal{V}_K\right)$.
\end{proposition}
\begin{proof}
Let $\Gamma := \{\gamma_j: j=1,2,...\}$ be a complete orthogonal basis system for $L^{2}(\Omega)$. 
Then by definition, for $u \in \mathcal{H}\subset L^{2}(\Omega\times\Omega)$, we have that $u(\omega_1, \omega_2) = \sum_{j=1}^\infty\sum_{l=1}^\infty a_{jl}\gamma_{j}(\omega_1)\gamma_{l}(\omega_2)$, where $a_{jl} = a_{lj}$, due to symmetry. Denote the infinite symmetric matrix 
$$
A = \begin{bmatrix} a_{11} & a_{12} & a_{13} & \cdots \\
a_{21} & a_{22} &  a_{23} & \cdots\\
a_{31} & a_{32} &  a_{33} & \cdots\\
\vdots & \vdots &  \vdots & \vdots\\
\end{bmatrix}.
$$
Since $u \in L^{2}(\Omega\times\Omega)$, it follows that $\sum_{j=1}^\infty\sum_{l=1}^\infty a_{jl}^2 < \infty$ and therefore $A$ determines a Hilbert–Schmidt operator and thus is compact. Then by the spectral theorem, there exists real eigenvalues $\{\lambda_k\}_{k=1}^\infty$ and orthogonal eigenvectors $\{e_k\}_{k=1}^\infty$ such that 
$$
A = \sum_{k=1}^\infty \lambda_ke_k\otimes e_k,
$$
\citepSupp{Hsing2015TheoreticalFO}. Let $\xi_k = \sum_{j=1}^\infty e_{kj}\gamma_j(\omega)$, it follows that both 
$$
\begin{aligned}
    u(\omega_1, \omega_2) &= \sum_{j=1}^\infty\sum_{l=1}^\infty a_{jl}\gamma_{j}(\omega_1)\gamma_{l}(\omega_2) \\
    &= \sum_{j=1}^\infty\sum_{l=1}^\infty\sum_{k=1}^\infty \lambda_ke_{kj} e_{kl}\gamma_{j}(\omega_1)\gamma_{l}(\omega_2) \\ 
    &= \sum_{k=1}^\infty\lambda_k\left(\sum_{j=1}^\infty e_{kj}\gamma_{j}(\omega_1)\right)\left(\sum_{l=l}^\infty e_{kl}\gamma_{l}(\omega_1)\right) \\ 
    &= \sum_{k=1}^\infty \lambda_k \xi_k(\omega_1)\xi_k(\omega_2)
\end{aligned}
$$
and 
$$
\begin{aligned}
\langle\xi_k,\xi_m\rangle_{L^{2}} &= \int_{\Omega}\left(\sum_{j=1}^\infty e_{kj}\gamma_{j}(\omega)\right)\left(\sum_{l=l}^\infty e_{jl}\gamma_{l}(\omega)\right)d\omega \\
&= \sum_{j=1}^\infty\sum_{l=l}^\infty e_{kj}e_{jl}\int_{\Omega}\gamma_{j}(\omega)\gamma_{l}(\omega)d\omega \\
&= \sum_{j=1}^\infty e_{kj}e_{mj} = \delta_{km},
\end{aligned}
$$
completing the proof. 
\end{proof}
Let $\{(\lambda_l^k,\xi_l^k)\}_{l}$ be the decomposition guaranteed by proposition~\ref{prop:uap} for eigenfunction $\psi_k$ and define  $\mathcal{V}_{\infty}^{*}=\{\xi_j^{*}\otimes\xi_j^{*}:j=1,2,...\}$ such that $\{\xi_j^{*}\}_{j=1}^\infty$ form a complete orthogonal basis system (CONS) in $L^{2}(\Omega)$. The summability condition in Assumption~\ref{assm:smoothness_class} can be written as:
$$
    \sum_{j=1}^\infty|\langle\sum_{l=1}^\infty \lambda_l^{k}\xi_{l}^k\otimes \xi_l^{k},\xi_{j}^*\otimes \xi_j^{*}\rangle_{L^{2}(\Omega\times\Omega)}| < \infty.
$$
We consider the upper bound given by:
\begin{equation}\label{eqn:summability_bound}
    \sum_{j=1}^\infty\sum_{l=1}^\infty |\lambda_l^{k}|\langle\xi_{l}^k, \xi_j^{*}\rangle_{L^{2}(\Omega)}^2 < \infty.
\end{equation}
Since $\{\xi_j^{*}\}_{j=1}^\infty$ is a CONS, we have that for all $l$ and $k$:
$$
\begin{aligned}
    \sum_{j=1}^\infty\langle\xi_l^{k},\xi_j^{*}\rangle_{L^{2}(\Omega)}^2 &= 
\int_{\Omega}\left(\sum_{j=1}^\infty\langle\xi_l^{k},\xi_j^{*}\rangle_{L^{2}(\Omega)}\xi_j^{*}(\omega)\right)^2d\omega \\ 
&= \|\xi_k^{l}\|_{L^{2}(\Omega)}^2 = 1,
\end{aligned}
$$
and thus Equation~\ref{eqn:summability_bound} reduces to the condition that $\sum_{l=1}^\infty |\lambda_l^{k}| < B_1$, for some $B_1<\infty$ and for all $k$.
\par 
To understand the implications of the absolute summability bound on the eigenvalues, we consider the following operator associated with each eigenfunction. As $\psi_k$ is symmetric and integrable, it defines the integral operator on $L^{2}(\Omega)$: $\Psi_k(u)(\omega) := \int_{\Omega} \psi_k(\omega, \omega^\prime ) u(\omega^\prime)d\omega^\prime$ for $u\in L^{2}(\Omega)$. It is easy to show that
$$
\|\Psi_k(u)\|_{L^{2}(\Omega)}^2\le  \|u\|_{L^{2}(\Omega)}^2\|\psi_k\|_{L^{2}(\Omega\times\Omega)}^2
$$
and thus $\Psi_k$ is bounded. A straight-forward consequence of the symmetry of $\psi_k$ is that $\Psi_k$ is self-adjoint. Furthermore, using the fact the $\Psi_k$ is compact and thus has a countable set of eigenvalues (4.2.3 \citeSupp{Hsing2015TheoreticalFO}), it is easy to show by definition that the eigenvalues of $\Psi_k$ are $\{\lambda_l^k\}_{l=1}^\infty$. Then Assumption~\ref{assm:smoothness_class} is equivalent  to the condition that all the $\Psi_k$'s are Trace class operators with $\text{Tr}(\Psi_k) < B_1$, as $\|\Psi_k\|_{Tr} = \sum_{l=1}^\infty |\lambda_{l}^k|$ for bounded linear self-adjoint operators \citepSupp{Hsing2015TheoreticalFO}.
\par 
We now consider what type of conditions on $\psi_k$ guarantees that $\text{Tr}(\Psi_k) < \infty$, for all $k$. In the context of familiar function classes, from theorem 13, ch. 30 in \citeSupp{lax2014functional}, a sufficient condition for $\Psi_k$ to be trace-class is $\psi_k$ being $\psi_k\in\mathcal{C}^{\infty}(\Omega\times\Omega)$. Though the uniform bound $\text{Tr}(\Psi_k) < B_1$ does not have as simple an interpretation in terms of familiar function classes, a characterization can be formed by assuming $\Psi_k$ is a composite operator of two bounded Hilbert-Schmidt operators:
\begin{proposition}\label{prop:composite_operator}
Assume $\exists$ symmetric $\psi_{k1},\psi_{k2}\in L^{2}(\Omega\times\Omega)$ such that
$$
\psi_k(\omega_1, \omega_2) = \int_{\Omega}\psi_{k1}(\omega_1,\omega^{*})\psi_{k2}(\omega^{*},\omega_2)d\omega^{*}
$$
and $\|\psi_{k1}\|_{\infty}\|\psi_{k2}\|_{\infty} < B_1$, then $\text{Tr}(\Psi_k) < B_1$.
\end{proposition}
\begin{proof}
Using the Cauchy–Schwarz inequality, the definition of the trace of an integral operator, and the unit measure assumption on $\Omega$, we have:
$$
\|\Psi_k\|_{Tr}\le \|\psi_{k1}\|_{L^{2}(\Omega\times\Omega)}\|\psi_{k2}\|_{L^{2}(\Omega\times\Omega)} \le \|\psi_{k1}\|_{\infty}\|\psi_{k2}\|_{\infty} < B_1
$$
\end{proof}
As a final point, we note that Assumption~\ref{assm:smoothness_class} can almost certainly be weakened, e.g., by using the ``interpolation spaces'' between $\mathcal{H}$ and $ L_{1,B_{1}}(\mathcal{V}_K^{*})$, as was the tactic in \citeSupp{barron2008} in their work on the approximation of unknown deterministic functions. We leave an extension of our results for random functions in this more general space as a direction of future work.

\section{Marginal Basis System}\label{sec:marg_splines}

\subsection{Spherical Splines}
Define a spherical triangle $T$ with vertices $\boldsymbol{v}_1, \boldsymbol{v}_2, \boldsymbol{v}_3$, as the set of points in $\mathbb{S}^2$ that are bounded by the great circle arcs connecting $(\boldsymbol{v}_{i}, \boldsymbol{v}_{i+1})$, $i=1,2,3$ and $\boldsymbol{v}_4\equiv \boldsymbol{v}_1$. A collection of spherical triangles $\mathcal{T} = \{T_j\}_{j=1}^J$ is a spherical triangulation of $\mathbb{S}^2$, provided that the intersection between any two elements in $\mathcal{T}$ is empty, a common vertex or a common edge and $\mathbb{S}^2 = \bigcup_{T_j\in\mathcal{T}} T_j$. The \textit{spherical Delaunay triangulation} is the unique triangulation over a set of points $\{\boldsymbol{v}_m\}_{m=1}^M$ such that the minimum angle in the triangulation is as large as possible.
\par 
For each spherical triangle $T_j$, we can define a local coordinate system called the \textit{spherical barycentric coordinates}, defined for $\boldsymbol{v}\in T_i$ to be the nonnegative real numbers satisfying 
$$
    \boldsymbol{v} = b_1(\boldsymbol{v})\boldsymbol{v}_1 + b_2(\boldsymbol{v})\boldsymbol{v}_3 + b_3(\boldsymbol{v})\boldsymbol{v}_3.
$$
We can define the linear spherical Bernstein basis on $T_j$ as the span of the three nodal basis functions $\mathcal{B} := \text{span}\left(\{b_1(\boldsymbol{v}), b_2(\boldsymbol{v}), b_3(\boldsymbol{v})\}\right)$. That is, for any $s(\boldsymbol{v}) \in \mathcal{B}$, there exists coefficients $c_1,c_2,c_3$ such that $s(\boldsymbol{v}) = \sum_{j=1}^3c_jb_j(\boldsymbol{v})$.
The local basis functions defined for each spherical triangle can be ``glued together'' to form a piecewise continuous linear spline space over $\mathbb{S}^2$. Following \citeSupp{lai2007}, we define 
$$
    \mathcal{S}_1^0(\mathcal{T}) = \{s\in C^0(\mathbb{S}^2):s|_{T_{j}}\in \mathcal{B},\quad\text{for}\quad j=1,...,J\}.
$$
An explicit basis for $ \mathcal{S}_1^0(\mathcal{T})$ can be constructed as follows: for each vertex $\boldsymbol{v}_m$ in tessellation $\mathcal{T}$, let $\phi_j\in\mathcal{S}_1^0(\mathcal{T})$ such that
$$
    \phi_j(\boldsymbol{v}_m) = \begin{cases}
      1 & \text{ if } j=m\\
      0 & \text{ otherwise}
    \end{cases}.
$$
It can be shown that $\phi_j$ is unique and that $\{\phi_j\}_{j=1}^M$ forms a basis for $\mathcal{S}_1^0(\mathcal{T})$ \citepSupp{lai2007}.
\par 
The $\phi_j$ have many attractive properties. In particular, efficient algorithms exist for evaluating and computing their directional derivative using the local basis functions. Additionally, their local support property is crucial for the development of our local inference procedure, outlined in Section~\ref{ssec:inference}. For further information, see the excellent references \citepSupp{lai2007,schumaker2015}. 

\subsection{Penalty Matrix Computation}\label{ssec:pen_matrix_construction}
\begin{proposition}\label{prop:penalty_formulation}
Under the basis expansion in Equation~\eqref{eqn:basis_expansion}, the quadratic variation penalty $\text{Pen}_{QV}(\xi_k) = \int_{\Omega}\|\nabla_{\Omega}\xi_k(\omega)\|^2d\omega$ has the representation 
$$
\begin{pmatrix} \boldsymbol{c}_{1} \\ \boldsymbol{c}_{2}  \end{pmatrix}^\intercal \begin{bmatrix} \int_{\mathbb{S}^2}\boldsymbol{F}_{1}\boldsymbol{F}^\intercal_{1} & \boldsymbol{0} \\ \boldsymbol{0} & \int_{\mathbb{S}^2}\boldsymbol{F}_{2}\boldsymbol{F}^\intercal_{2} \end{bmatrix} \begin{pmatrix} \boldsymbol{c}_{1} \\ \boldsymbol{c}_{2}  \end{pmatrix},
$$
where $\boldsymbol{F}_d(\boldsymbol{x})\in\mathbb{R}^{M_{d}\times 3}$ is the Jacobian of $\boldsymbol{\phi}_{M_{d}}$.
\end{proposition}
\begin{proof}
Since $\Omega = \mathbb{S}_1^2 \cup \mathbb{S}_2^2$, the gradient separates as 
$$
    \nabla_{\Omega}(\xi_k(\boldsymbol{x})) = \nabla_{\mathbb{S}^2}(\boldsymbol{c}^\intercal_{1,k}\boldsymbol{\phi}_{M_{1}}(\boldsymbol{x}))\mathbb{I}\{\boldsymbol{x}\in\mathbb{S}_1^2\} + \nabla_{\mathbb{S}^2}(\boldsymbol{c}^\intercal_{2,k}\boldsymbol{\phi}_{M_{2}}(\boldsymbol{x}))\mathbb{I}\{\boldsymbol{x}\in\mathbb{S}_2^2\}.
$$
We have 
$$
\begin{aligned}
    \text{Pen}(\xi_k) &=  \int_{\Omega}\left[ \boldsymbol{c}^\intercal_{1,k}\boldsymbol{F}_{1}(\boldsymbol{x})\boldsymbol{F}^\intercal_{1}(\boldsymbol{x})\boldsymbol{c}_{1,k}1\{\boldsymbol{x}\in\mathbb{S}_1^2\} + \boldsymbol{c}^\intercal_{2,k}\boldsymbol{F}_{2}(\boldsymbol{x})\boldsymbol{F}^\intercal_{2}(\boldsymbol{x})\boldsymbol{c}_{2,k}1\{\boldsymbol{x}\in\mathbb{S}_2^2\}\right] d\boldsymbol{x} \\
    &=  \begin{pmatrix} \boldsymbol{c}_{1,k} \\ \boldsymbol{c}_{2,k} \end{pmatrix}^\intercal \begin{bmatrix} \int_{\mathbb{S}^2}\boldsymbol{F}_{1}\boldsymbol{F}^\intercal_{1} & \boldsymbol{0} \\ \boldsymbol{0} & \int_{\mathbb{S}^2}\boldsymbol{F}_{2}\boldsymbol{F}^\intercal_{2} \end{bmatrix} \begin{pmatrix} \boldsymbol{c}_{1,k} \\ \boldsymbol{c}_{2,k} \end{pmatrix} \\ 
\end{aligned}
$$
\end{proof}
Fast algorithms exist for computing the required directional derivatives to form $\nabla_{\mathbb{S}^2}$ \citepSupp{lai2007}. The required integrals can be computed using Lebedev quadrature rules or using exact rules for the local planar triangle approximation to the spherical triangle. The latter of these is of course faster and preferable when the spherical triangles are small enough.
\subsection{Selecting the Triangulation}\label{ssec:triangulation_selection}

It is well established that the geometric properties of the spherical triangles in $\mathcal{T}_{d}$ and the spatial distribution of vertices can heavily affect the approximation power of $\boldsymbol{\phi}_{M_{d}}$ \citepSupp{shewchuk2002}. It is, therefore, important to have a procedure to construct $\mathcal{T}_d$. Considering the complex geometry of the white surface, the grid points $\boldsymbol{X}$ are usually irregularly spaced over $\Omega$. We use a simple pruning heuristic to design $\mathcal{T}_d$, independently for each copy of $\mathbb{S}^2_d$, such that the local density of basis functions is aligned to the spatial distribution of $\boldsymbol{X}$. In particular, our approach consists of the following steps: i) initiate a dense set of (nearly) equispaced vertices over $\mathbb{S}^2$, ii) compute the pairwise distances between this vertex set and the grid $\boldsymbol{X}$, iii) sequentially prune vertices that have the largest minimum distance to the grid points until $M_d$ vertices are left, and iv) define $\mathcal{T}_d$ to be the spherical Delaunay tessellation \citepSupp{lai2007} over this refined vertex set. This simple heuristic was observed to consistently result in better $L^2$ reconstruction performance than the (nearly) equispaced vertex strategy for the same $M_d$. Figure~\ref{fig:grid_and_tess} shows the tessellaton resulting from this pruning method that was used to define the marginal spline basis system in the real data experiments on both hemispheres along with the high-resolution grid points $\boldsymbol{X}$. 

\begin{figure}[!ht]
    \centering
    \includegraphics[width=\textwidth]{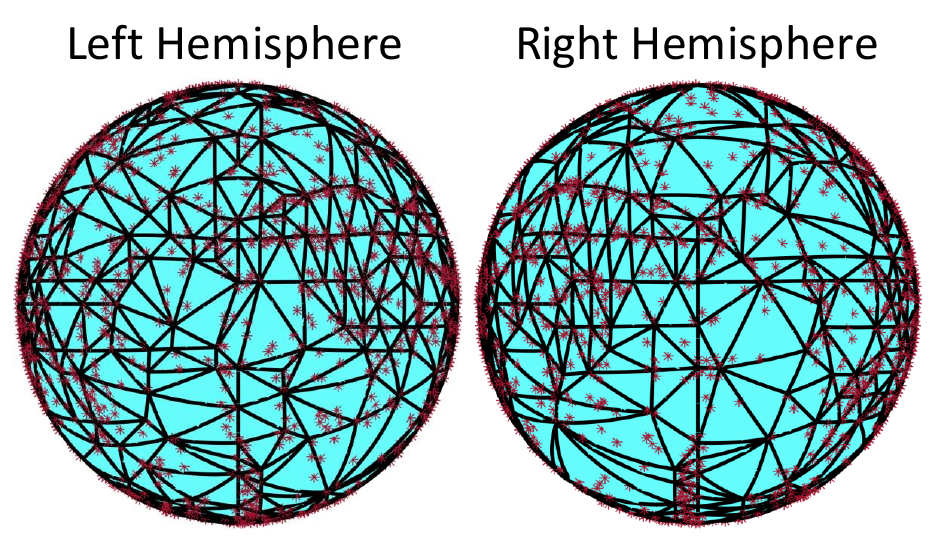}
    \caption[Grid points $\boldsymbol{X}$ (red stars) and corresponding spherical tessellation $\mathcal{T}_1$,$\mathcal{T}_2$ for $M_1=M_2=410$.]{Grid points $\boldsymbol{X}$ (red stars) and corresponding spherical tessellation $\mathcal{T}_1$,$\mathcal{T}_2$ for $M_1=M_2=410$.}
    \label{fig:grid_and_tess}
\end{figure}
\newpage
\section{Algorithm}\label{ssec:algoImp}
\par\bigskip
\noindent{\textbf{Derivation of Approximation in Equation~\eqref{eqn:discrete_l2_norm}}}
\par\bigskip

The approximation in Equation~\eqref{eqn:discrete_l2_norm} uses the matrix element-wise $l^2$ inner product (e.g., $\left<,\right>_F$) as a discrete approximation to the continuous integral in the $L^2(\Omega\times\Omega)$ inner product, under the assumption that the grid points $\boldsymbol{X}$ are relatively dense and reasonably spaced over $\Omega$. To see why this is the case, consider the following. Denote the disjoint partition formed by the spherical Voronoi tesselation of $\boldsymbol{X}$: $\bigcup_i V_i = \Omega$ with $x_i$ being the center of $V_i$ \citepSupp{du2003}. Using Riemann sums, the $L^2(\Omega\times\Omega)$ inner product of functions $u_{1}:\Omega\times\Omega\mapsto\mathbb{R}$,  $u_{2}:\Omega\times\Omega\mapsto\mathbb{R}$ can be approximated as 
$$
\int_{\Omega\times\Omega}u_1(\omega_1,\omega_2)u_2(\omega_1,\omega_2)d\mu(\omega_1)d\mu(\omega_2)\approx \sum_{i}\sum_{j}u_1(x_i,x_j)u_2(x_i,x_j)\mu(V_i)\mu(V_j) 
$$
for surface measure $\mu$, as the partition becomes arbitrarily fine. Assuming that the element measures are reasonably similar to one another, i.e.  $\mu(V_i)\approx\text{Const.}$ $\forall i$, (the points $x_i$ are well dispersed and dense), we can make the further approximation
\begin{equation}\label{eqn:voronoi_approx}
    \int_{\Omega\times\Omega}u_1(\omega_1,\omega_2)u_2(\omega_1,\omega_2)d\mu(\omega_1)d\mu(\omega_2)\approx \text{Const.}\sum_{i,j}u_1(x_i,x_j)u_2(x_i,x_j).
\end{equation}
The desired result follows after noting that the right-hand side of Equation~\ref{eqn:voronoi_approx} is proportional to the $\left<,\right>_F$ of the evaluation matrices of $u_1$ and $u_2$ on $\boldsymbol{X}\times\boldsymbol{X}$.
\par\bigskip
\noindent{\textbf{Derivation of Block Updates}}
\par\bigskip 
We first state and prove two results that will be useful for our derivation.
\begin{proposition}\label{prop:isometric_mapping}
Define the reparameterization $\tilde{\boldsymbol{c}} = \boldsymbol{D}\boldsymbol{V}^\intercal\boldsymbol{c}$. Then 
\begin{equation}
\sum_{i=1}^N\left\langle \boldsymbol{R}_{k-1,i}, \boldsymbol{\Phi}\boldsymbol{c}\otimes\boldsymbol{\Phi}\boldsymbol{c}\right\rangle_{F}^2 \propto \sum_{i=1}^N\left(\tilde{\boldsymbol{c}}^\intercal s_{i}(\tilde{\boldsymbol{c}})\boldsymbol{G}_{k-1,i}\tilde{\boldsymbol{c}}\right)
\end{equation}
where $\boldsymbol{G}_{k-1,i} = \boldsymbol{U}^\intercal\boldsymbol{R}_{k-1,i}\boldsymbol{U}$ and $s_i(\tilde{\boldsymbol{c}}) = \tilde{\boldsymbol{c}}^\intercal\boldsymbol{G}_{k-1,i}\tilde{\boldsymbol{c}}$. 
\end{proposition}
\begin{proof}
Notice that
$$
\begin{aligned}
\left\langle \boldsymbol{R}_{k-1,i}, \boldsymbol{\Phi}\boldsymbol{c}\otimes\boldsymbol{\Phi}\boldsymbol{c}\right\rangle_{F} &= \text{trace}\left( \boldsymbol{R}_{k-1,i}^\intercal, \right[\boldsymbol{\Phi}\boldsymbol{c}\left]\right[\boldsymbol{\Phi}\boldsymbol{c}\left]^\intercal\right)\\ 
&= \text{trace}\left( \boldsymbol{R}_{k-1,i}^\intercal, \boldsymbol{U}\boldsymbol{D}\boldsymbol{V}^\intercal\boldsymbol{c}\boldsymbol{c}^\intercal\boldsymbol{V}\boldsymbol{D}\boldsymbol{U}^\intercal\right) \\
&= \tilde{\boldsymbol{c}}^\intercal\boldsymbol{G}_{k-1,i}\tilde{\boldsymbol{c}}.
\end{aligned}
$$
Coupled with Equation~\eqref{eqn:coef_rewrite}, the result follows directly. 
\end{proof}
For clarity, in the following derivation we work mainly in the transformed parameter $\boldsymbol{\tilde{c}}$. For notational convenience, we define the quantities 
 $$
 \boldsymbol{\tilde{J}}_{\boldsymbol{\phi}} = \boldsymbol{D}^{-1}\boldsymbol{V}^\intercal\boldsymbol{J}_{\boldsymbol{\phi}}\boldsymbol{V}\boldsymbol{D}^{-1}, \qquad \boldsymbol{\tilde{Q}}_{\boldsymbol{\phi}} = \boldsymbol{D}^{-1}\boldsymbol{V}^\intercal\boldsymbol{Q}_{\boldsymbol{\phi}}\boldsymbol{V}\boldsymbol{D}^{-1},
$$ 
so
$$
    \boldsymbol{c}^\intercal\boldsymbol{J}_{\boldsymbol{\phi}}\boldsymbol{c} = \boldsymbol{\tilde{c}}^\intercal\boldsymbol{\tilde{J}}_{\boldsymbol{\phi}}\boldsymbol{\tilde{c}}, \qquad
    \boldsymbol{c}^\intercal\boldsymbol{Q}_{\boldsymbol{\phi}}\boldsymbol{c} = \boldsymbol{\tilde{c}}^\intercal\boldsymbol{\tilde{Q}}_{\boldsymbol{\phi}}\boldsymbol{\tilde{c}}.
$$

\begin{proposition}\label{prop:orthogonal_constraint}
Let $\boldsymbol{\tilde{C}}_{k-1} = [\boldsymbol{\tilde{c}}_1, ..., \boldsymbol{\tilde{c}}_{k-1}]$. Then the orthogonality constraint $\boldsymbol{c}^\intercal\boldsymbol{J}_{\boldsymbol{\phi}}\boldsymbol{c}_j = \boldsymbol{\tilde{c}}^\intercal\boldsymbol{\tilde{J}}_{\boldsymbol{\phi}}\boldsymbol{\tilde{c}_j} = 0\text{ for }j=1,...,k-1$ is equivalent the condition that 
$$
    (\boldsymbol{I} - \boldsymbol{\tilde{P}}_{k-1})\boldsymbol{\tilde{c}} = \boldsymbol{\tilde{c}}
$$
where 
$$
\begin{aligned}
    \boldsymbol{\tilde{P}}_{k-1} &= \boldsymbol{\tilde{C}}_{k-1}[\boldsymbol{\tilde{C}}_{k-1}^\intercal\boldsymbol{\tilde{J}}_{\boldsymbol{\phi}}\boldsymbol{\tilde{C}}_{k-1}]^{-1}\boldsymbol{\tilde{C}}_{k-1}^\intercal\boldsymbol{\tilde{J}}_{\boldsymbol{\phi}} \\
    &= \boldsymbol{D}\boldsymbol{V}^\intercal\boldsymbol{C}_{k-1}\left[\boldsymbol{C}_{k-1}^{\intercal}\boldsymbol{J}_{\boldsymbol{\phi}}\boldsymbol{C}_{k-1}\right]^{-1}\boldsymbol{C}_{k-1}^{\intercal}\boldsymbol{J}_{\boldsymbol{\phi}}\boldsymbol{V}\boldsymbol{D}^{-1} \\
    &= \boldsymbol{P}_{k-1} 
\end{aligned}
$$
\end{proposition}
\begin{proof}
Note that the matrix $\boldsymbol{\tilde{J}}_{\boldsymbol{\phi}}$ is symmetric and positive definite and therefore defines an inner product on $\mathbb{R}^{M_{1}+M_{2}}$. The result then follows trivially from standard results in linear algebra. 
\end{proof}

Using Proposition~\ref{prop:isometric_mapping}, we can reparamterize problem \eqref{eqn:empirical_optm_reformulated_discretized_AO} as 
\begin{equation}\label{eqn:empirical_optm_reformulated_discretized_reparam} 
\begin{aligned}
    \hat{\tilde{\boldsymbol{c}}}_k = &\underset{\tilde{\boldsymbol{c}} \in \mathbb{R}^{M}}{\text{ argmax }} \sum_{i=1}^N\left(\tilde{\boldsymbol{c}}^\intercal \boldsymbol{s}_{i}\boldsymbol{G}_{k-1,i}\tilde{\boldsymbol{c}}\right) - \alpha_1\boldsymbol{\tilde{c}}^\intercal\boldsymbol{\tilde{Q}}_{\boldsymbol{\phi}}\boldsymbol{\tilde{c}}\\
    \textrm{s.t.} \quad & \boldsymbol{\tilde{c}}^\intercal\boldsymbol{\tilde{J}}_{\boldsymbol{\phi}}\boldsymbol{\tilde{c}} = 1,  \quad \boldsymbol{\tilde{c}}^\intercal\boldsymbol{\tilde{J}}_{\boldsymbol{\phi}}\boldsymbol{\tilde{c}}_j  = 0 \text{ for }j=1,2,...,k-1\\
    \quad & \left\|\boldsymbol{V}\boldsymbol{D}^{-1}\tilde{\boldsymbol{c}}\right\|_{0} \le n_{\alpha_{2}} \\
    \quad & \boldsymbol{s} = \mathcal{G}_{k-1}\times_1 \boldsymbol{\tilde{c}} \times_2 \boldsymbol{\tilde{c}},
\end{aligned}
\end{equation}
where the rewritten constraint $\boldsymbol{s} = \mathcal{G}_{k-1}\times_1 \boldsymbol{\tilde{c}} \times_2 \boldsymbol{\tilde{c}}$ follows from the element-wise definition of $\boldsymbol{s}_i(\boldsymbol{\tilde{c}})$ in Proposition~\ref{prop:isometric_mapping} coupled with the definition of $d$-mode multiplication. Noting that we have $\mathcal{G}_{k-1}\times_3\boldsymbol{s} = \sum_{i=1}^N\boldsymbol{s}_{i}\boldsymbol{G}_{k-1,i}$, \eqref{eqn:empirical_optm_reformulated_discretized_reparam} becomes 
$$
\begin{aligned}
    \hat{\tilde{\boldsymbol{c}}}_k = &\underset{\tilde{\boldsymbol{c}} \in \mathbb{R}^{M}}{\text{ argmax }} \tilde{\boldsymbol{c}}^\intercal\left[\mathcal{G}_{k-1}\times_3\boldsymbol{s} - \alpha_1{\boldsymbol{\tilde{R}}}_{\boldsymbol{\phi}}\right]\boldsymbol{\tilde{c}}\\
    \textrm{s.t.} \quad & \boldsymbol{\tilde{c}}^\intercal\boldsymbol{\tilde{J}}_{\boldsymbol{\phi}}\boldsymbol{\tilde{c}} = 1,  \quad \boldsymbol{\tilde{c}}^\intercal\boldsymbol{\tilde{J}}_{\boldsymbol{\phi}}\boldsymbol{\tilde{c}}_j  = 0 \text{ for }j=1,2,...,k-1\\
    \quad & \left\|\boldsymbol{V}\boldsymbol{D}^{-1}\tilde{\boldsymbol{c}}\right\|_{0} \le n_{\alpha_{2}}  \\
    \quad & \boldsymbol{s} = \mathcal{G}_{k-1}\times_1 \boldsymbol{\tilde{c}} \times_2 \boldsymbol{\tilde{c}} 
\end{aligned}
$$
Applying Proposition~\ref{prop:orthogonal_constraint}, the orthogonality constraint can be absorbed into the objective function, resulting in 
$$
\begin{aligned}
    \hat{\tilde{\boldsymbol{c}}}_k = &\underset{\tilde{\boldsymbol{c}} \in \mathbb{R}^{M}}{\text{ argmax }} \tilde{\boldsymbol{c}}^\intercal\left(\boldsymbol{I} - \boldsymbol{\tilde{P}}_{k-1}\right)\left[\mathcal{G}_{k-1}\times_3\boldsymbol{s} - \alpha_1{\boldsymbol{\tilde{R}}}_{\boldsymbol{\phi}}\right]\left(\boldsymbol{I} - \boldsymbol{\tilde{P}}_{k-1}\right)\boldsymbol{\tilde{c}}\\
    \textrm{s.t.} \quad & \boldsymbol{\tilde{c}}^\intercal\boldsymbol{\tilde{J}}_{\boldsymbol{\phi}}\boldsymbol{\tilde{c}} = 1 \\
    \quad & \left\|\boldsymbol{V}\boldsymbol{D}^{-1}\tilde{\boldsymbol{c}}\right\|_{0} \le n_{\alpha_{2}} \\
    \quad & \boldsymbol{s} = \mathcal{G}_{k-1}\times_1 \boldsymbol{\tilde{c}} \times_2 \boldsymbol{\tilde{c}}. 
\end{aligned}
$$
Transforming back to the original coordinates using $\boldsymbol{c} = \boldsymbol{V}\boldsymbol{D}^{-1}\tilde{\boldsymbol{c}}$ and invoking the relaxed norm constraint, 
$$
\begin{aligned}
     \hat{\boldsymbol{c}}_k  = \underset{\boldsymbol{c} \in \mathbb{R}^{M}}{\text{ argmax }} &\quad \boldsymbol{c}^\intercal\left[\boldsymbol{V}\boldsymbol{D}\left(\boldsymbol{I} - \boldsymbol{P}_{k-1}\right)\left[\mathcal{G}_{k-1}\times_3\boldsymbol{s} - \alpha_1\boldsymbol{D}^{-1}\boldsymbol{V}^\intercal\boldsymbol{Q}_{\boldsymbol{\phi}}\boldsymbol{V}\boldsymbol{D}^{-1}\right]\left(\boldsymbol{I} - \boldsymbol{P}_{k-1}^{\intercal}\right)\boldsymbol{D}\boldsymbol{V}^\intercal\right]\boldsymbol{c} \\
    & \textrm{s.t.} \quad \boldsymbol{c}^\intercal\boldsymbol{c} = 1, \quad \left\|\boldsymbol{c}\right\|_{0} \le n_{\alpha}, \\
     \quad & \boldsymbol{s} = \mathcal{G}_{k-1}\times_1 \left(\boldsymbol{D}\boldsymbol{V}^\intercal\boldsymbol{c}\right) \times_2 \left(\boldsymbol{D}\boldsymbol{V}^\intercal\boldsymbol{c}\right), 
\end{aligned} 
$$
and the updates \eqref{eqn:ctilde_update} and \eqref{eqn:s_update} follow trivially by applying block coordinate ascent to the above. 
\par\bigskip 
\noindent{\textbf{Initialization}}
\par\bigskip 

The block update \eqref{eqn:ctilde_update} requires an initialization for $\boldsymbol{c}^{(0)}$ and $\boldsymbol{s}^{(0)}$. Notice that, under the rank-1 approximation of $\mathcal{Y}$, we have that 
$$
    \mathcal{G} = \mathcal{Y}\times_1 \boldsymbol{U}^{\intercal}\times_2\boldsymbol{U}^\intercal
    \approx \left(\boldsymbol{\Phi}\boldsymbol{c} \otimes \boldsymbol{\Phi}\boldsymbol{c}  \otimes \boldsymbol{s} \right)\times_1 \boldsymbol{U}^\intercal\times_2\boldsymbol{U}^\intercal
    =  \left(\boldsymbol{D}\boldsymbol{V}^{\intercal}\boldsymbol{c} \otimes \boldsymbol{D}\boldsymbol{V}^\intercal\boldsymbol{c}  \otimes\boldsymbol{s} \right).
$$
Then by properties of the $d$-mode matricization, 
\begin{equation}\label{eqn:krp_prop}
\mathcal{G}_{(1)} \approx \boldsymbol{D}\boldsymbol{V}^\intercal\boldsymbol{c}\left(\boldsymbol{D}\boldsymbol{V}^\intercal\boldsymbol{c} \odot \boldsymbol{s}\right)^\intercal, \quad \mathcal{G}_{(3)} \approx \boldsymbol{s}\left(\boldsymbol{D}\boldsymbol{V}^\intercal\boldsymbol{c} \odot \boldsymbol{D}\boldsymbol{V}^\intercal\boldsymbol{c}\right)^\intercal,
\end{equation}
where $\odot$ denotes the Kronecker product and $_{(d)}$ denotes the $d$-mode matricization. Equation~\eqref{eqn:krp_prop} indicates that, for the $k$'th rank-1 factor, the leading left singular vectors of $\mathcal{G}_{k-1,(3)}$ and, under the transform $\boldsymbol{V}\boldsymbol{D}^{-1}$, $\mathcal{G}_{k-1,(1)}$ are a reasonable initialization for the block variables $\boldsymbol{s}^{(0)}$ and $\boldsymbol{c}^{(0)}$, respectively. In practice, these can be computed efficiently using power iterations on the symmetric matrices $\mathcal{G}_{k-1,(1)}\mathcal{G}^{\intercal}_{k-1,(1)}$ and $\mathcal{G}_{k-1,(3)}\mathcal{G}^{\intercal}_{k-1,(3)}$. 

\section{Additional Results}\label{sec:additional_results}
\subsection{Brain Network and Traits}\label{ssec:brain_network_traits}
Figure~\ref{fig:global_inference_Desikan} shows the results of the hypothesis testing task described in Section~\ref{sssec:hypothesis_testing} and Table~\ref{tab:desikan_prediction_results} shows the results from the prediction task from Section~\ref{sssec:trait_prediction}, where the discrete network-based approaches were applied to the connectomes formed via the Desikan atlas. Coupled with the results displayed in Figure~\ref{fig:global_inference} and Table~\ref{tab:concon_vs_atlas}, we make the following two important observations: i) our method (CC) again strongly outperforms the competitors on the Desikan atlas for most outcomes ii) the results from the competitor methods are highly variable between atlases.
\begin{figure}[t]
    \centering
    \includegraphics[scale=0.58]{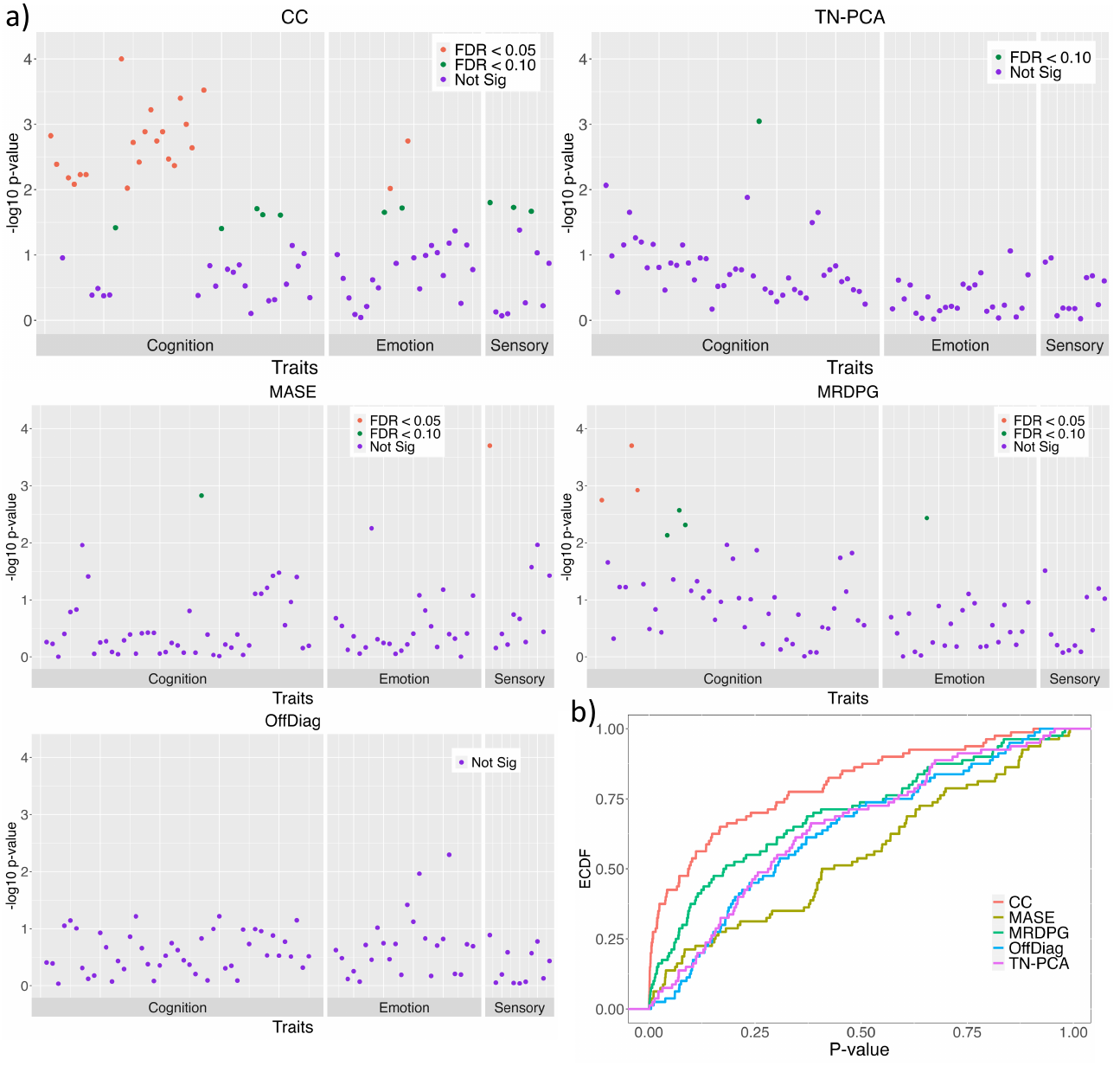}
    \caption{Results from the global hypothesis testing task using the Desikan atlas. a) $-\text{log}_{10}$-transformed uncorrected p-values from the MMD test for all 80 traits. b) Empirical cumulative distribution functions (ECDFs) of the p-values for each analysis method.}
    \label{fig:global_inference_Desikan}
\end{figure}

\begin{table}[htbp]
\centering
\tiny
\begin{tabular}{r|rrrr}
\toprule
 &   \multicolumn{4}{c}{\textbf{Embedding}}  \\
Outcome & TNPCA & MASE & MRDP & OffDiag \\
\midrule
Fluid Intelligence-1 & 0.1789 (0.0074) & -0.1025 (0.0090) & -0.0046 (0.0097) & 0.0898 (0.0083) \\
Self Regulation-1 & 0.0938 (0.0081) & -0.0094 (0.0088) & -0.0105 (0.0088) & 0.1586 (0.0077) \\
Self Regulation-3 & 0.1510 (0.0093) & -0.0061 (0.0094) & -0.0795 (0.0077) & -0.0244 (0.0103) \\
Sustained Attention-2 & 0.0355 (0.0088) & -0.0585 (0.0084) & -0.0825 (0.0093) & 0.0229 (0.0083) \\
Executive Function-1 & 0.0002 (0.0086) & 0.0431 (0.0091) & 0.0699 (0.0087) & 0.1086 (0.0091) \\
Executive Function-2 & -0.0459 (0.0086) & 0.0070 (0.0085) & 0.0066 (0.0083) & -0.0717 (0.0107) \\
Psychiatric-3 & 0.0508 (0.0095) & -0.0569 (0.0078) & -0.0176 (0.0094) & 0.1052 (0.0096) \\
Psychiatric-6 & -0.0315 (0.0077) & 0.1068 (0.0101) & -0.0235 (0.0093) & 0.1604 (0.0094) \\
Taste-1 & -0.0878 (0.0087) & 0.1155 (0.0082) & 0.0396 (0.0083) & -0.0054 (0.0093) \\
Emotion Recognition-1 & 0.0307 (0.0092) & 0.0938 (0.0092) & 0.0182 (0.0086) & 0.0559 (0.0085) \\
Emotion Recognition-2 & 0.0136 (0.0083) & -0.0518 (0.0071) & 0.0118 (0.0093) & 0.0775 (0.0103) \\
Emotion Recognition-4 & 0.1113 (0.0104) & -0.0594 (0.0086) & -0.0042 (0.0084) & 0.0292 (0.0108) \\
Anger and Hostility-2 & -0.0301 (0.0083) & -0.0683 (0.0086) & -0.0351 (0.0092) & 0.0574 (0.0091) \\
Anger and Hostility-3 & -0.0820 (0.0073) & 0.0400 (0.0071) & 0.1181 (0.0092) & -0.0032 (0.0091) \\
BMI and Weight-1 & 0.2082 (0.0095) & 0.0462 (0.0102) & 0.0647 (0.0088) & 0.2420 (0.0093) \\
\hline
\end{tabular}
    \caption{Pearson correlations of the trait prediction task for TN-PCA, MASE, MRDPG, and OffDiag using the Desikan atlas.}
\label{tab:desikan_prediction_results}
\end{table}

\subsection{Reproducibility Analysis}\label{ssec:reproducibility}
\subsubsection{Scan-Rescan Reproducibility}
We are interested in studying the reproducibility of the proposed reduced-rank representation across scans. That is, we want to understand how well the reduced-rank function space, $\text{span}\left(\mathcal{V}_K\right)$, estimated from one population, can represent the re-scanned data from the same population. For this evaluation, we utilize the HCP scan-rescan data, which consists of a subset of HCP subjects for whom two scans were collected during different sessions several months apart. This dataset contains 37 subjects and a total of 74 scans. In the absence of disease, we expect the brain organization of a healthy adult to be relatively constant over the span of several months, and therefore the HCP scan-rescan data can be considered as independent noisy samples of the same connectome, making it appealing for evaluating reproducibility. In particular, we used the scan-1 data to form estimates $\hat{\xi}_k$ according to the procedure in Section~\ref{sec:methods}. For several $K$, the estimated $\mathcal{V}_K = \{\hat{\xi}_1, ...,\hat{\xi}_K\}$ were used to represent the scan-2 data by standard basis expansion under the least squares principle. In Figure~\ref{fig:reproducibility_analysis}a, we show the first three coefficients corresponding to $\{\hat{\xi}_1,\hat{\xi}_2,\hat{\xi}_3\}$ for a random collection of 10 test-retest subjects. The unique markers identify multiple scans from the same individual. One can see that a within-subject clustering pattern is already present using only $K=3$ basis functions. 
\par
The pairwise $L^2(\Omega\times\Omega)$ distances between the basis expansion representations of all the scan-1 and scan-2 data were computed. Figure~\ref{fig:reproducibility_analysis}b displays the pairwise distance matrices for several different $K$'s. Each matrix is $74 \times 74$, giving the pairwise distances between the $74$ scans from $37$ subjects. Scans from the same subject are put next to each other, generating small $2\times2$ blocks along the diagonal. A quantitative assessment of the reproducibility was obtained using the leave-one-out cross-validated (LOOCV) accuracy of the nearest neighbor classifier applied to the distances: for each representation $i$, we compute $j = \text{argmin}_{l\neq i} \text{dist}(i,l)$, where $\text{dist}(i,l)$ is shorthand for the distance between scans $i$ and $l$. The LOOCV accuracy is computed as the fraction of examples where $i$ and $j$ are two scans corresponding to the same individual. At the bottom of Figure~\ref{fig:reproducibility_analysis}b, we show the LOOCV accuracy results. With $K=22$, we achieved a $100\%$ classification accuracy. 
\begin{figure}
    \centering
   \includegraphics[scale=0.43]{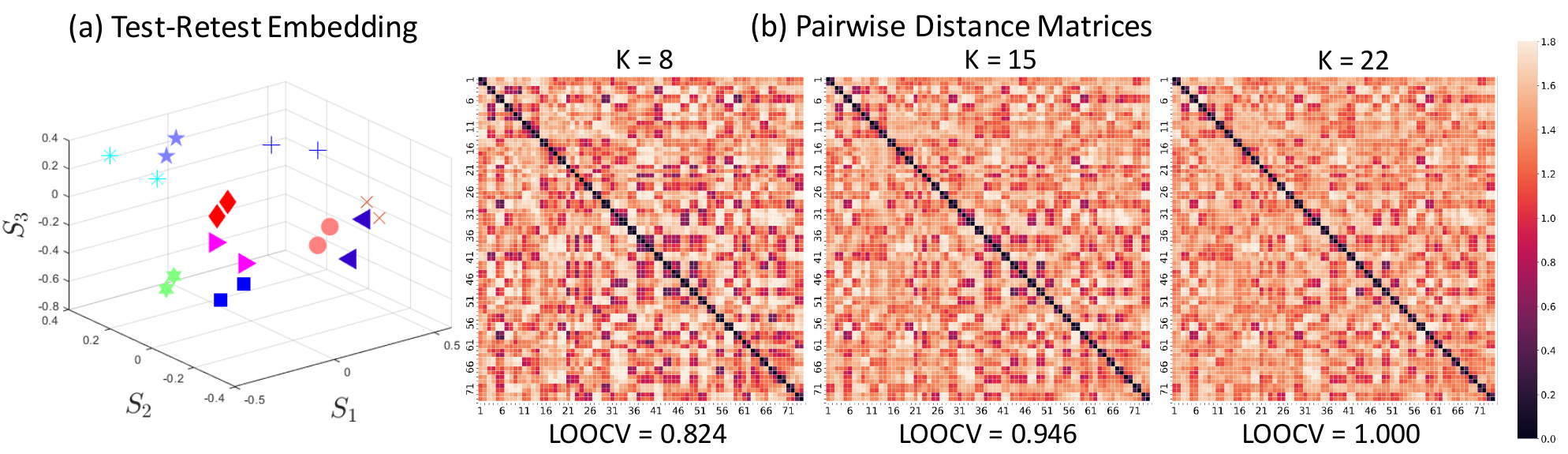}
    \caption{(a) The low-rank continuous embeddings for 10 randomly selected HCP test-retest subjects using $K=3$. Each subject is identified by a unique marker. (b) Pairwise distance matrices for $74$ scans from $37$ subjects under several $K$'s. The LOOCV accuracy of the nearest neighbor classier applied to the distances is also recorded.}
    \label{fig:reproducibility_analysis}
\end{figure}
\subsection{Reproducibility Across Grid Resolutions}\label{sec:high_res_vs_low_res}
\begin{figure}[!ht]
    \centering
    \includegraphics[scale=0.5]{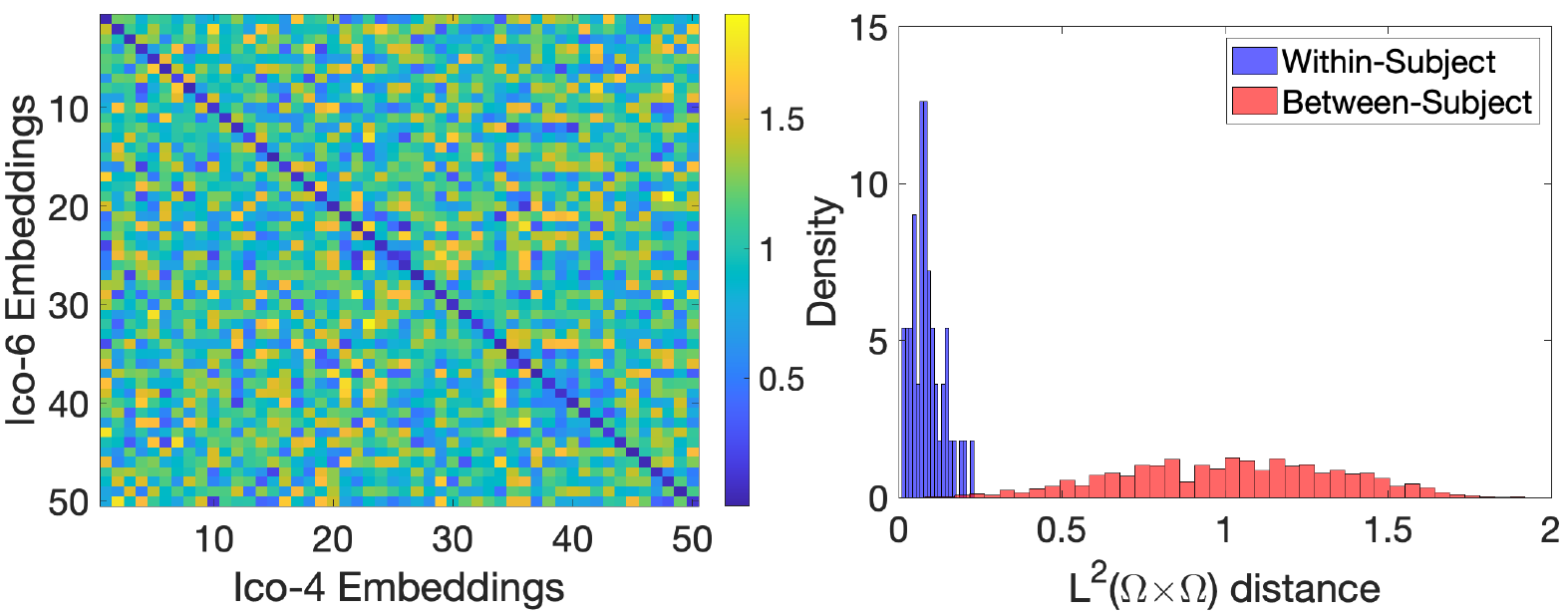}
    \caption{(left) Pairwise distance matrix of the continuous connectome emdeddings in the space spanned by $\mathcal{V}_{K}^{(1)}$ for both icospherical grids. Distances between emdeddings for the same subject are along the diagonal. (Right) The corresponding distribution of within subject distances (diagonal) and between subject distances (off diagonal) from the pairwise distance matrix on the left.}
    \label{fig:hr_lr}
\end{figure}
A primary advantage of the proposed methodology is its agnosticism to the grid $\boldsymbol{X}$. Hence, another form of reproducibility of interest is between grid resolutions. In other words, we are interested in how representations computed from the connectivity data of the same subject at multiple different grid resolutions compare to each other. Therefore, in this section, we compare the reproducibility of the continuous embeddings estimated from connectivity data at multiple resolutions.
\par 
Although storage considerations were the primary reason for the initial dimension reduction of the HCP connectome data from $64,000^2$ to $4,121^2$ vertices, the point-wise smoother using the heat-kernel from \citeSupp{moyer2017}, i.e. Equation~\eqref{eqn:KDE_cc}, itself can become a computational bottleneck for extremely dense grids. As discussed in Section~\ref{ssec:subj_level_estim}, there are alternative estimators for the point-wise smoothed connectivity that can easily be swapped into our framework, and fast point-wise smoothing for high-dimensional connectivity data is an active area of research \citeSupp{borovitskiy2021}. To avoid this computational bottlekneck in this analysis, we use a simple fast alternative local smoother, defined as follows:
\begin{enumerate}
    \item Triangulate the grid points $\boldsymbol{X}$ on $\Omega$.
    \item For each pair of grid points  $(\omega_1,\omega_2)\in\boldsymbol{X}\times \boldsymbol{X}$, obtain each pair of triangles which have $\omega_1$ or $\omega_2$ as a vertex. Denote these sets of triangles as $T_{\omega_{1}}$ and $T_{\omega_{2}}$, respectively.
    \item For any endpoint $p_1^{j},p_2^{j}$ which falls within a pair of triangles in $(\tau_1,\tau_2)\in T_{\omega_{1}}\times T_{\omega_{2}}$, assign weight $w^{j}(\omega_1,\omega_2)$ to be the product of the barycentric coordinates of $p_1^{j}$ in $\tau_1$ with respect to the $\omega_1$ and  $p_2^{j}$ in $\tau_2$ with respect to the $\omega_2$.
    \item Form
    $\widehat{U}(\omega_1,\omega_2)$ as 
    the sum of the $w^{j}(\omega_1,\omega_2)$ over all endpoints. 
\end{enumerate}
This simple endpoint smoothing method has been used in the literature and will soon be available in the SBCI pipeline \citepSupp{cole2021}. 
\par 
In this analysis, we use data from 
the Adolescent Brain Cognitive Development (ABCD) study \citepSupp{casey2018abcd}. Full imaging and acquisition protocol for this data can be found in the same reference. We randomly select 50 baseline (year 1) scans and run them through the same image processing pipeline as outlined in Section~\ref{ssec:image_processing}, except we use the smoother defined above to form point-estimates at two resolutions, 5,124 total grid points and 81,924 total grid points, formed via an icosphere parcellation with subdivision 4 and 6, respectively. We estimate a data-driven basis from the connectivity data at both resolutions, denoted $\mathcal{V}_{K}^{(4)}$ for the $5,124$ grid point and $\mathcal{V}_{K}^{(6)}$ for the $81,924$ grid points,  using the same model and algorithm parameters outlined in Section~\ref{sec:real_data_experiments} with $K=10$. We then project the representations in $\mathcal{V}_{K}^{(6)}$ to the space spanned by $\mathcal{V}_{K}^{(4)}$ and compute the pairwise distances between all functions.
\par 
Figure~\ref{fig:hr_lr} (left) shows the resulting pairwise distance matrix, where the rows and columns are organized so that the same subject's data at different resolutions appears along the diagonal. The right plot of Figure~\ref{fig:hr_lr} shows the corresponding within subject distances, i.e. the diagonal, in blue and the between subject distances in red. We see clear separation between subjects, despite using only $K=10$ basis functions and forming the pointwise estimates $\widehat{U}(\omega_1, \omega_2)$ using the rather crude estimator outlined above. This shows both that our method is scalable to extremely high dimensional grids and produces consistent and reproducible function representations between grid-size granularities.

\subsection{Comparison with Super-High Resolution Atlas-Based Method} \label{supp:highres}
We conducted an investigation comparing our method with high-resolution atlases using data from 300 ABCD subjects, each with two scans, one from baseline and a 1-year follow-up. Specifically, we experimented with a parcellation encompassing approximately 4,000 small triangular parcels. Raw fiber count between each pair of triangular parcels was computed as the discrete connectivity. Our findings indicate that raw high-resolution atlases significantly reduce the reproducibility of our connectivity results and are not effective predictors of cognitive traits. In particular, the atlas-based connectivity, when evaluated using dICC (analogous to ICC for quantifying reproducibility, as discussed in greater detail in \citeSupp{zhang2018mapping}), yields a value less than 0.55. In contrast, the CC's dICC value is over 0.75. Similarly, CC's predictive power for behavioral traits (e.g., picture vocabulary age-adjusted and crystallized intelligence age-adjusted) significantly surpasses that of the high-resolution atlas-based connectivity.  The correlation between CC predicted picture vocabulary measures and the actual measures is 0.31 while the value for the raw high-resolution connectivity is 0.27. The values for the crystallized trait are 0.32 and 0.28, respectively.
\par 
We attribute these results to two key factors. First, the network derived from the high-resolution atlas is exceedingly sparse, making network estimation unreliable. For instance, while we typically construct about $10^6$ streamlines or fiber curves, an atlas with 4000 parcels can yield up to $10^7$ connections, and most pairs of parcels are unconnected. Second, the presence of small nodes complicates node alignment, leading to inflated variance among connections due to the node misalignment issue (nodes in classical brain network analysis are presumed to be matched).

\bibliographystyleSupp{chicago}
\bibliographySupp{references}

\end{document}